\documentstyle[aps,epsf,prc,floats,twocolumn]{revtex}
\begin{document}
%
%
\draft
\title{Double Beta Decays of $^{100}$Mo and $^{150}$Nd }
\author{A.~De~Silva, M.~K.~Moe, M.~A.~Nelson, and M.~A.~Vient}
\address{Department of Physics \& Astronomy \\
University of California, Irvine, CA 92697}
\date{\today}     
\maketitle
\begin{abstract}
The double beta decays of $^{100}$Mo and $^{150}$Nd were studied in a Time
Projection Chamber located 72 m underground.
A 3275~h exposure of a 16.7~g sample of metallic
Mo enriched to 97.4\% in $^{100}$Mo resulted in a two-neutrino
half-life, $T_{1/2}^{2\nu}\!=\!(6.82^{+0.38}_{-0.53}\pm 0.68)\!\times\!
10^{18}$~y.
Similarly, a 6287~h exposure of 15.5~g of
Nd$_{2}$O$_{3}$ enriched to 91\% in $^{150}$Nd yielded 
$T_{1/2}^{2\nu}\!=\!(6.75^{+0.37}_{-0.42}\pm 
0.68)\!\times\! 10^{18}$~y.
Lower limits on half-lives for neutrinoless decay with 
and without majoron emission also have been measured.

\end{abstract}
%
%
%

\section{Introduction}

\vspace{4mm}

Double beta decay, the single isobaric jump of two 
units in atomic number, 
\[
\rm (A,Z) \longrightarrow (A,Z+2) + 2 e^{-} + 2 \bar{\nu}_{e}\;,
\]
is a rare second-order weak transition,
directly measurable under favorable circumstances by means 
of the ejected pair of $\beta$ particles.  
The phenomenon has been observed, with its two accompanying 
neutrinos ($\beta\beta_{2\nu}$), in several nuclei for which 
single beta decay is strongly inhibited or energetically 
forbidden~\cite{moevogel}.  
The search for a variation 
lacking the usual pair of neutrinos in the final state, is underway in 
several laboratories as
a uniquely sensitive probe of the mass and charge conjugation properties of the 
neutrino.  Should a neutrinoless double beta decay ($\beta\beta_{0\nu}$) 
branch be seen, it would 
demonstrate that at least one neutrino is a massive Majorana 
particle~\cite{kayser}, at variance with the lepton 
conserving, massless Dirac particles of the standard model.

A second nonstandard mode sometimes considered is neutrinoless decay 
with majoron emission ($\beta\beta_{0\nu,\chi}$).  The majoron is a 
hypothetical boson, coupling to the neutrino with sufficient strength in 
recent models, to make a significant contribution to the $\beta\beta$-decay
rate~\cite{berezhiani,burgess1,burgess2,burgess3}.

The three decay modes are distinguishable experimentally by the 
spectrum of the sum of the two $\beta$-particle energies.  
The $\beta\beta_{0\nu}$ mode is characterized by a distinctive line 
spectrum at Q$_{\beta\beta}$, 
whereas the $\beta\beta_{2\nu}$ electron 
sum spectrum is a broad distribution, peaking at about 1/3 of the 
Q$_{\beta\beta}$ value.
The various proposed $\beta\beta_{0\nu,\chi}$
spectra are also 
broad distributions, generally distinct from $\beta\beta_{2\nu}$.

While all $\beta\beta$ decays observed by direct counting 
experiments have had the features of standard $\beta\beta_{2\nu}$, 
stringent lower limits on $\beta\beta_{0\nu}$ and 
$\beta\beta_{0\nu,\chi}$ half-lives have been achieved.
Geochemical and radiochemical $\beta\beta$ experiments, and 
searches for $\gamma$-rays following $\beta\beta$ decay to excited 
levels of the daughter,  
do not distinguish among decay modes, but when their rates have been 
compared with direct-counting limits on the exotic modes,
a predominance of $\beta\beta_{2\nu}$ has always been implied~\cite{ijmp}. 

The extraction of upper limits on neutrino mass and neutrino-majoron 
coupling strength (or of actual values for these parameters, should 
the $\beta\beta_{0\nu}$ and $\beta\beta_{0\nu,\chi}$ half-lives ever 
be measured) depends on complicated nuclear matrix element 
calculations.  Comparison of theoretical and experimental half-lives 
for the observed $\beta\beta_{2\nu}$ mode is an important test 
for these calculations~\cite{moevogel}.  

Here we describe our final measurements~\cite{nelson} of two relatively fast 
$\beta\beta$ emitters, $^{100}$Mo (Q$_{\beta\beta} = 3034\pm 6$~keV) 
and $^{150}$Nd (Q$_{\beta\beta} = 3367.1\pm 2.2$~keV)~\cite{wapstra}.  
We also discuss the disappearance of a persistent 
excess of
high-energy $\beta\beta_{0\nu,\chi}$-like events
that were present in our earlier data.


\section{Experimental Apparatus}
\label{aparatus}

\subsection{The TPC}

The measurements were carried out in a time projection chamber (TPC) 
located in an underground valve house in a canyon wall at 
the Hoover Dam.  This site provided a minimum of
72~m of rock shielding, and reduced the cosmic ray muon flux 
by a factor of approximately 130 from that at the Earth's
surface.  

\begin{figure}
\mbox{\epsfxsize=8.6 cm\epsffile{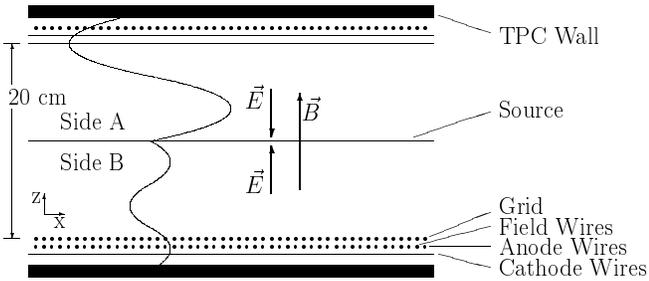}}
\vskip 8mm
\caption{Schematic representation of the source plane
and wire array geometry.  The sinusoids represent the x-z view of 
helical tracks from two electrons emitted by the source.}
\label{tpcgeom}
\end{figure}

The TPC has been described in detail 
in Refs.~\cite{nim,erice}.  It is a rectangular polycarbonate (Lexan) 
box of inside dimensions 88~cm $\times$ 88~cm $\times$ 23~cm.
The $\beta\beta$ source plane bisects the volume into two 10-cm-deep drift
regions, and serves as the central drift field electrode.  Planes of
sensing wires parallel to the source are located near
the TPC walls.  Anode and cathode wires are mutually 
perpendicular, and provide $x$ and $y$ coordinates.  The $z$ 
coordinate is provided by the arrival time of drifting ionization 
electrons at the anode.
Spatial resolution in $x$ and $y$ is 
determined by the wire spacing, 5.1~mm. 
The 5.0~mm $z$-resolution is established by the 5~mm/$\mu$s 
drift velocity and 1~MHz frequency
of wire read-out.  The corresponding 1~$\mu$s ``time buckets'' are clocked into an 
80-deep shift register, in which the full 10~cm drift distance spans 
the first 20 buckets. 
The remaining 60 buckets are used to
record any hits which occur in the 1~ms following the initial trigger. 
This scheme permits tagging background from  $^{214}$Bi events 
by recording the subsequent 164-$\mu$s $^{214}$Po
$\alpha$-particle.
The required drift 
field is established by a negative potential applied 
to the source plane.  
Drift field and anode wire potentials are compensated continuously for 
fluctuations in barometric pressure.
The geometry of the
source and wires is illustrated in Fig.~\ref{tpcgeom}.

The TPC is shielded on all sides by 15~cm of lead.  Notoriously radioactive
materials, such as circuit board, solder, ribbon
cable, circuit components,  and connectors, are all completely outside 
of the lead shield.  
Interleaved with the lead shielding are six cosmic-ray 
veto panels in a $4\pi$ arrangement 
which identifies muon associated events 
at a rate of approximately 0.25~Hz.

A magnetic field is applied 
perpendicular to the source plane by a pair of 
coils mounted in an iron flux return.
An electron emitted from the
source follows
a helical trajectory  about the $z$-axis.
The momentum of the particle and its angle to the magnetic field 
are determined from
the parameters of the helix fitted to its track.
The 
1200 gauss field (approximately twice the strength in our earlier 
experiments without the flux return)
is uniform to $\pm 0.8$\% over the active volume of the TPC.

Helium
mixed with 7\% propane 
flows through the chamber at a rate of 57~l/hr.  Mixture
proportions and flow rate are both controlled by a mass-flow
controller system.  Prior to mixing, 
the helium is passed through a liquid-nitrogen
cold trap, which freezes out all measurable Rn produced by 
Ra within the gas supply cylinders.  

The trigger used in the present experiment 
requires a start hit and at least one hit in
buckets 3-5.  In addition to complete tracks, the trigger will
capture tracks having a $z$ component as short as three 
time buckets.  This feature 
increases the efficiency at which short $\alpha$-particle tracks
are saved, while maintaining some protection from
noisy wires.  Here we refer to $\alpha$-particles other than the 
$^{214}$Po type already included in 1~ms sensitive period
following a $^{214}$Bi trigger.
An $\alpha$-particle appearing within an hour or two at the site of a 
$\beta\beta$ candidate event is a good indicator that the event was 
background from one of the primordial decay chains.
The trigger fires 
at about 0.6~Hz.  

A small dead time in the chamber arises primarily 
from the cosmic ray veto.
Triggers are vetoed for a period of 30~$\mu$s following a hit in any
of the veto panels.  
The resulting dead time is about 0.6\%.  
The overall dead time of the system is
$(0.87\pm 0.05)$\%.

The data
acquisition system is able to direct the latch modules to disable
the occasional noisy wire.  Typically, the TPC is operated with about 1\% of
the wires dead or disabled.

Routines for off-line selection of one- and two-electron events are 
discussed in Sec.~\ref{analysis}.  Only those 2e$^{-}$ pairs with 
electrons emerging from opposite sides of the source are analyzed.

\subsection{Isotope Preparation}

The molybdenum, enriched to 97.4\% in $^{100}$Mo,  was purchased as MoO$_{3}$ 
from a German supplier, and is of Russian origin.  The oxide is not the 
optimum form for detection of  $\alpha$-particles associated with background
processes.  As an example, the 7.7~MeV $^{214}$Po
$\alpha$-particle range is diminished from about 22~mg/cm$^{2}$ 
in the metal to about 10~mg/cm$^{2}$
of Mo in the oxide.
For this reason, the MoO$_{3}$ was reduced to the metal in our
laboratory in a quartz tube in a high temperature hydrogen atmosphere.
About 20.0~g of fine metallic powder were available for source fabrication.

The $^{150}$Nd consisted of 3~g of
Nd$_{2}$O$_{3}$ purchased from the same German 
supplier, plus 
$\approx 1/4$ of an 82~g batch of
Russian Nd$_{2}$O$_{3}$ obtained through an
agreement with the late Professor A.~A.~Pomansky of the
Institute for Nuclear Research in Moscow.  Elemental Nd is 
reactive, so it was left as the oxide.

Samples of both isotopes were forwarded to Pacific 
Northwest Laboratories (PNL) for photon counting and mass spectroscopy 
(ICPMS).  The PNL results indicated radioisotopes of 
europium in the $^{150}$Nd at
levels too great for $\beta\beta$-decay studies. 
The INR and commercial batches of neodymium oxide then 
were sent to the Ames Laboratory Materials Preparation Center, where
they were combined and processed with a liquid chromatography system.
The $^{150}$Nd isotopic enrichment of the combined batches was 90.9\%.
Of the initial 85~g
of Nd$_{2}$O$_{3}$, 94\% survived the purification process.
Samples were again 
forwarded to PNL where post-purification measurements indicated a 
reduction of more than two orders of magnitude in europium 
contamination~\cite{brodzinski}.  This reduction was more than 
sufficient, although 
we were requested not to reveal the 
level of sensitivity of the measurements.
Uranium and Thorium were not 
detected.

Meanwhile our portion of the Nd$_{2}$O$_{3}$, which had been stored 
in sealed plastic
bags, changed color from its characteristic robin's-egg blue, to a
blue-grey.  This indicated that
some of the oxide had been hydrated to Nd(OH)$_{3}$. 
This compound is undesirable for the same reasons discussed in
the case of the $^{100}$Mo oxide.  
The hydrate was transferred
to a quartz crucible and heated to 1100 $^{\circ}$C in an electric
furnace.  This restored the characteristic blue color, 
and the approximately 12\% loss in mass observed is consistent with
conversion of Nd(OH)$_{3}$ to
Nd$_{2}$O$_{3}$.  After processing, 
approximately 19.0~g remained.
It was stored in an evacuated jar until
needed.  

\subsection{Source Preparation}

A thin layer of $^{100}$Mo was
deposited on each of two 4~$\mu$m aluminized polyester 
(Mylar) substrates, each supported by 
a half-section of a Lexan source frame assembly.  
A nitrogen gas powered gun
was used to inject the powdered isotope material at high speed into a
glass box positioned in turn over each substrate.  The
resulting cloud of powder settled onto the Mylar, leaving a
thin, uniform deposit of isotope.  This fragile
layer of powder was then fixed to the Mylar with a fine misting of
polyvinyl acetal resin (Formvar) solution.  
A 1~pCi droplet of $^{207}$Bi solution was applied to the geometrical
center of one of the source halves in a 3 cm square region left devoid 
of powder by a mask.  After drying, the Bi was fixed
in place with a drop of Formvar solution 
to prevent migration to the $\beta\beta$ deposit.
(The conversion electrons emitted by the $^{207}$Bi 
provide a continuous monitor of detector performance.)
When dry, the two substrates were
placed face-to-face to form a sandwich of Mylar and isotope, with the
isotope in the interior, and the aluminized surface of the Mylar 
facing outward.

All seams between the
Mylar and the Lexan frame were sealed with low activity 
epoxy, isolating the interior of the source from the TPC gas.
The interior of the source was vented to atmosphere through an oil 
bubbler, allowing the slight
overpressure of the TPC to press the sheets of source Mylar tightly together
into a thin, flat plane.
Both the $^{100}$Mo and $^{150}$Nd $\beta\beta$ sources were prepared in
the same manner.  
The masses of the isotope deposits on the two sources were
$(16.7\pm 0.1)$~g of Mo and
$(15.5 \pm 0.1)$~g of Nd$_{2}$O$_{3}$.
Full details of the source preparation procedure may be
found in Ref.~\cite{nelson}.


\section{Detector Performance}
\label{perform}

\subsection{$\alpha$-Particle Detection Probability}
\label{pap}

Detection of  $\alpha$-particles is an important aid in 
identification of $\beta\beta$-decay backgrounds.  The thicknesses of
the two isotope deposits were chosen to assure that a large fraction of
$\alpha$-particles produced on or in the $\beta\beta$ source would escape
to the chamber gas where they would be observed.
The probability P$_{\alpha}$ that an alpha particle from the 
source will enter the TPC gas has been calculated with the
CERN Library GEANT Monte Carlo, applied to the geometry and 
composition of the source.  P$_{\alpha}$ was also measured,
for the case of the $^{214}$Po $\alpha$ particle,
following an injection of $^{222}$Rn of sufficient strength to overwhelm 
pre-existing activity. 
Here P$_{\alpha}(^{214}$Po) was taken as the number of single 
electrons of energy $>$ 1.5~MeV (essentially all were $^{214}$Bi $\beta$ 
particles) that had an alpha track in the following 
millisecond, divided by the 
sum with or without an alpha.  For details, see Ref.~\cite{nelson}.  
The $^{214}$Bi from the Rn 
injection settles on the source surface, but as long as the source 
thickness is less than the alpha particle range, P$_{\alpha}$ 
determined in this way is a good approximation for $^{214}$Bi 
located anywhere on or within the source.
The results of these methods are given in 
Table~\ref{aranges}.

To the extent that $^{214}$Bi is on the surface of the source plane 
($\approx$100\% for the Mo source and $\approx$50\% for the Nd source,
Sec.~\ref{useries}),
the $^{214}$Po $\alpha$-decay can be identified
with higher efficiency than implied by these escape probabilities.
For surface $^{214}$Bi, even the 
$\alpha$-particle that buries itself completely in the source, is nearly always 
accompanied by 
the release of 
``shake-off'' electrons, which produce a one- or two-bucket
``blip'' at the $x$,$y$ location of the decay.  Observation of such a blip in
the 1 ms post-trigger interval identifies the trigger event as a
$^{214}$Bi decay.  This association is confirmed by the delay-time 
distribution of
blips, which shows the characteristic 164~$\mu$s half-life of
$^{214}$Po.  

Since the blips substantially increase
the identification chances for the $^{214}$Po $\alpha$ particle (see 
Table~\ref{aranges})
they 
are used as well as $\alpha$-particle
tracks in the routine rejection of $^{214}$Bi. 
The alpha detection
enhancement from shake-off electrons on the $^{150}$Nd source is smaller  
because of subsurface $^{226}$Ra present in the $^{150}$Nd$_{2}$O$_{3}$.
Other $\alpha$-decays used for
background estimation are 
$^{212}$Bi and $^{212}$Po, also included in   
Table~\ref{aranges}.

\begin{table*}
\caption[]{Ranges and escape probabilities for several important 
$\alpha$-particles.  
Values in parenthesis
for $^{214}$Po refer to $\alpha$-decay identification probability when
``blips'' from shake-off electrons are used in addition to full 
$\alpha$-particle tracks.
Energies are taken from 
Kaplan~\cite{kaplan}. Measured P$_{\alpha}$ values from Ref.~\cite{nelson} 
have been corrected for loss of $^{214}$Bi events when 
the $\alpha$ comes within $\sim 15$ $\mu$s and spoils the $\beta$ track, 
or comes after the 1024 $\mu$s window.}   
\begin{tabular}{cr@{${}\pm{}$}lr@{${}\pm{}$}lr@{${}\pm{}$}lcr@{${}\pm{}$}lc}
& \multicolumn{2}{c}{} & \multicolumn{2}{c}{}
& \multicolumn{6}{c}{Estimated Escape Probability, P$_{\alpha}$ (\%)} \\
& \multicolumn{2}{c}{$\alpha$ Energy} & \multicolumn{2}{c}{Mean Range} 
& \multicolumn{3}{c}{$^{100}$Mo source} & \multicolumn{3}{c}{$^{150}$Nd source} \\
$\alpha$-Decay & \multicolumn{2}{c}{(MeV)} & \multicolumn{2}{c}{in air (cm)}
& \multicolumn{2}{c}{Measured} & GEANT & \multicolumn{2}{c}{Measured} & GEANT \\
\tableline \rule{0mm}{2.75ex} 
$^{214}$Po & 7.6804 & 0.0009 & 6.907 & 0.006 &   
     73 & 5 & 75 & 72 & 5 & 76 \\
& \multicolumn{2}{c}{} & \multicolumn{2}{c}{} & 
	(98 & 2) & & (85 & 7) & \\
$^{212}$Bi & 6.0466 & 0.0027 & 4.730 & 0.008 & 60 & 7 \tablenotemark[1]
     & 65 & 58 & 7 \tablenotemark[1] & 66 \\
$^{212}$Po & 8.7801 & 0.0040 & 8.570 & 0.007 & 79 & 4 \tablenotemark[1]
     & 79 & 78 & 4 \tablenotemark[1] & 80 \\
\end{tabular} 
\tablenotetext[1]{Scaled from the $^{214}$Po measurement by the 
Bragg-Kleeman approximation for $\alpha$-particle ranges in 
materials.}
\label{aranges}
\end{table*}

\subsection{Opening Angle Acceptance}
\label{opangacc}

The TPC's inability to reconstruct tracks making angles to the 
magnetic field with cosines near 0~or~1, 
the requirement that
electrons from a $\beta\beta$-decay emerge from opposite sides of the
source, and multiple scattering of the electrons within the source, 
all conspire to distort the measured opening angle distribution and 
obliterate the distinction between opening angle distributions for 
$\beta\beta$ decay and background.  Because attempts to salvage the 
distributions through Monte Carlo corrections have not been very 
successful, we do not make use of opening angle information in our 
present analysis.  

However, it is worth pointing out that the requirement for 
opposite-side electrons leads to low efficiency for
small opening angles, and thereby helps suppress one of the potential
$\beta\beta$-decay  backgrounds, M\"{o}ller scattering (see
Sec.~\ref{backgrounds}).  M\"{o}ller opening angles are in the range 
$\cos\vartheta_{\rm M}\!\approx\! +0.2$
to $+0.5$.  The efficiency for
reconstructing events in this range is relatively 
low, and will selectively suppress M\"{o}ller events.

\subsection{Energy Resolution}
\label{ic}
                                                                              
A convenient line source for displaying the energy resolution of the TPC
is $^{207}$Bi, which provides internal conversion (IC) energies 
at approximately 0.5, 1.0, and 1.7~MeV.
Although the dozen conversion
electrons per hour emitted from the 1~pCi $^{207}$Bi deposit in the
center of the $\beta\beta$-decay source is sufficient for day-to-day
monitoring of detector performance, a stronger source was required 
temporarily for
accurate energy resolution studies.  This is particularly true for high
energies, since the 1.7~MeV $^{207}$Bi conversion line is extremely
weak.  Therefore, a separate source assembly was built and peppered 
with 16 deposits of $^{207}$Bi totalling 6~nCi, distributed over 
the area normally covered with powdered $\beta\beta$-decay isotope. 

\begin{figure}
\vskip -0.5 cm
\mbox{\epsfxsize=8.6 cm\epsffile{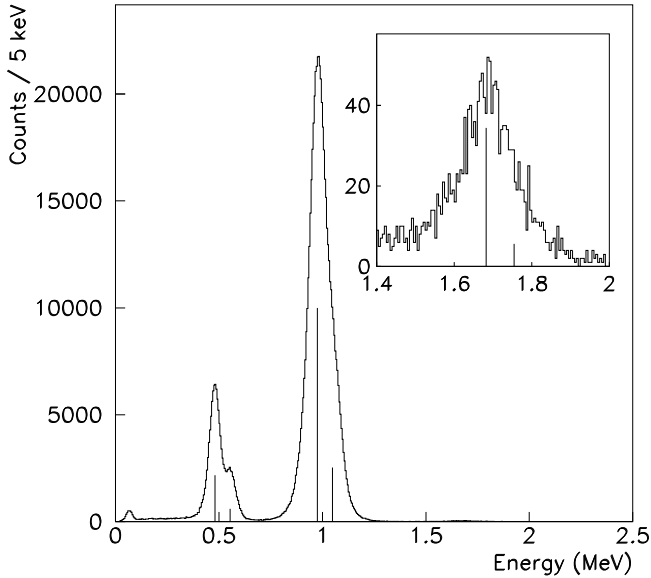}}
\caption{The measured energy spectrum of
$7\!\times\! 10^5$ $^{207}$Bi IC electrons.  Inset figure shows the weak peak
near 1.7~MeV.  Electrons with tracks nearly parallel to the magnetic
field lines ($|\cos \vartheta | > 0.8$) have been cut.  Conversion line
energies and relative strengths are represented by vertical lines.}
\label{bispec}
\end{figure}

\begin{figure}
\vskip -0.5 cm
\mbox{\epsfxsize=8.6 cm\epsffile{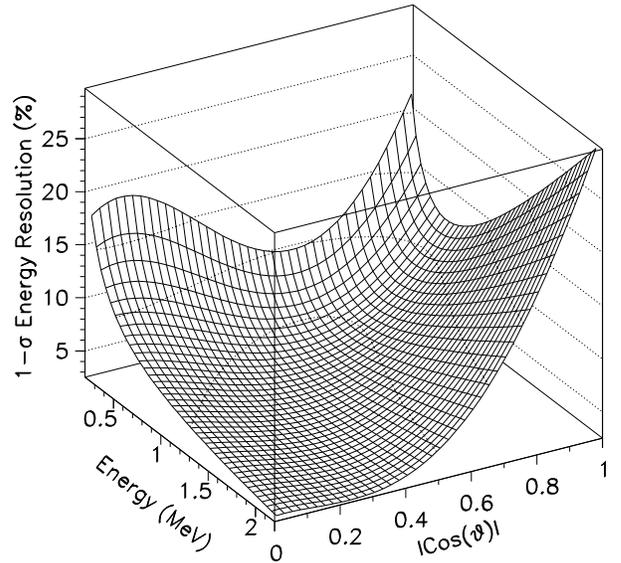}}
\caption{Energy resolution, in percent, of the
TPC as a function of electron energy and the cosine of the angle that
electron makes with the magnetic field lines.  This function is
derived from studying the conversion lines of $^{207}$Bi.}
\label{ecoserr}
\end{figure}

The resulting energy spectrum of $7\!\times\!10^{5}$ events 
is shown in Fig.~\ref{bispec}.  Energies of electrons with tracks
nearly parallel to the magnetic field ($|\cos \vartheta | > 0.8$)
are less well determined, and have been excluded from this
spectrum.  The widths of the peaks are determined by fitting each of
them with a pair of Gaussians, centered on the $K$ and $L$ lines.  
The corresponding $\sigma$ is 26~keV, 39~keV and 68~keV for the
482~keV, 976~keV and 1682~keV $K$ lines, respectively.  
The width of each
conversion line can be measured at various cosines, and a model of
resolution versus cosine and energy can be constructed. 
Figure~\ref{ecoserr} shows the results of this analysis.  This
function provides a statistical estimate of the uncertainty in the
energy of a given electron track. The Monte Carlo program, described in 
Sec.~\ref{response}, uses this function
to randomly smear the energies of the simulated events.

\subsection{Energy Threshold}
\label{energythreshold}

The energy of a particle is determined by fitting a helix to the
three-dimensional reconstruction of its trajectory.  
Low-momentum particles will create helical tracks of relatively small
radius and low pitch.  As the diameter of the helix approaches the
$x$- and $y$-resolution of the TPC, the quality of the reconstruction 
diminishes.
Similarly, if the $z$-separation of adjacent cycles of the helix is
less than the $z$-resolution of the TPC, the helix will not be
resolved, and only a single, wide swath of ionization will be
reconstructed. 

Although some sensitivity clearly remains for energies 
as low as the 57~keV 
$^{207}$Bi Auger electron~\cite{erice},
for this analysis a 250~keV energy threshold was imposed.  The TPC
does not reach optimum efficiency until somewhat higher energies,
but in place of still higher energy cuts, which would
sacrifice a significant amount of good data, a model of the detector
response was determined and used in subsequent analyses.

Also included in the model is the high-energy response 
which is limited as well, because the pitch
and radius of the track tend to increase with higher energies. 
At some point, the track becomes too straight for the fitter to
determine reliable helix parameters.
Fortunately, a large fraction of events satisfy the requirements for
good fits, and can be very nicely reconstructed.  
                       
\subsection{Detector Response Model}
\label{response}

One could determine the detector response to a particular $\beta$ 
spectrum by placing a calibrated $\beta$ source in the TPC, and noting
the ratio of the measured spectrum to the true spectrum 
as a function of energy.
Such a function would not be 
generally applicable, however, because its dependence on the
energy derivatives 
cannot be 
extracted from a single test spectrum.  For this reason, it is useful 
to separate the detector effects into two classes --- intrinsic and 
instrumental, and deal with the former by Monte Carlo.  The intrinsic 
effects include scattering and 
energy loss of electrons within the $\beta\beta$ source, and 
backscattering of electrons from the TPC walls.  Instrumental effects 
begin with the collection of ionization electrons from the 
gas, and include such factors as the sampling cell size, electron 
attachment by impurities, gas amplification, the readout electronics, 
and track reconstruction software.  

Intrinsic effects were modeled with the 
GEANT code.  
The model could then 
be applied generally 
for any $\beta$ spectrum.  More importantly, the model can 
predict {\it two}-electron behavior where the electrons are correlated in 
energy and opening angle, as for example 
$\beta\beta$ decay and 
M\"{o}ller scattering.

Instrumental effects are too complicated for a reliable Monte Carlo.
Their two consequences are the resolution function, already determined 
from $^{207}$Bi measurements in Sec.~\ref{ic}, and 
the probability that an electron that has already survived the GEANT process 
for intrinsic effects will also survive the track reconstruction process.
Since this probability is the same for any electron emerging from the source 
with a given 
energy and polar angle, it can be deduced from
a comparison of a TPC-measured $\beta$ spectrum with the resolution-smeared
output of the 
corresponding intrinsic-effects Monte Carlo.
This was done with 
the equilibrium $^{90}$Sr, $^{90}$Y combination $\beta$ spectrum.

\begin{figure}
\vskip -0.5 cm
\mbox{\epsfxsize=8.6 cm\epsffile{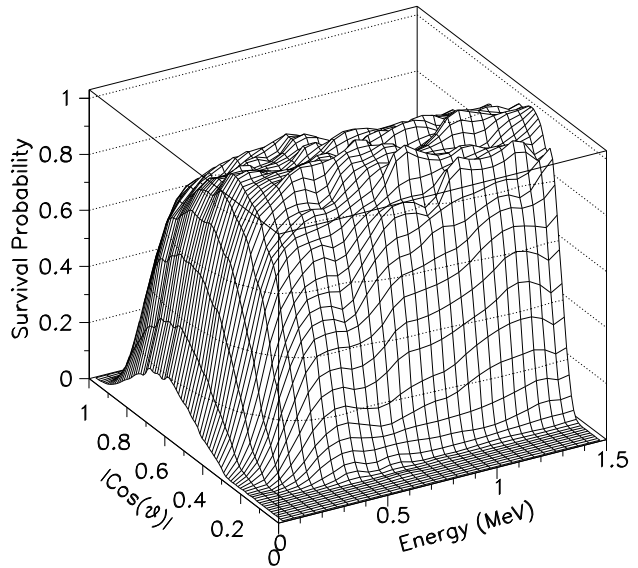}}
\caption{The normalized detector response 
as a function of energy and $\cos(\vartheta$), the 
angle between the track and the magnetic field.  
This is the electron survival probability after its emergence from the source 
plane.
}
\label{instru}
\end{figure}

\begin{figure}
\vskip -0.5 cm
\mbox{\epsfxsize=8.6 cm\epsffile{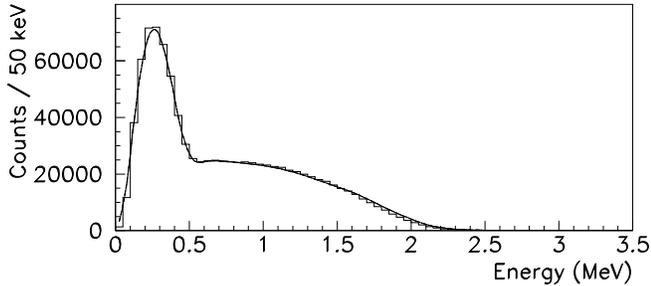}}
\caption{The measured $^{90}$Sr lone electron spectrum (histogram) 
and the simulated spectrum (solid line.)  The simulation includes energy 
smearing and the detector response model.}
\label{sr90lone}
\end{figure}

Input electrons for the Monte Carlo were sampled 
from the $^{90}$Sr and $^{90}$Y spectra of 
Ref.~\cite{betaspec}.  The output energies then were smeared by the 
resolution function of Sec.~\ref{ic}.  Comparison with the spectrum of 
reconstructed electrons from a calibrated drop of $^{90}$Sr solution 
applied to  a dummy $\beta\beta$ source in the TPC, revealed the 
survival probability 
of an electron after emerging from the source plane (Fig.~\ref{instru}.)  
This survival probability, extrapolated for energies above 1.5~MeV, 
was then included in the GEANT Monte Carlo and used to regenerate the 
$^{90}$Sr lone electron spectrum; the results compared with 
the corresponding measured spectrum are shown in Fig.~\ref{sr90lone}.  

\begin{figure}
\vskip -0.5 cm
\mbox{\epsfxsize=8.6 cm\epsffile{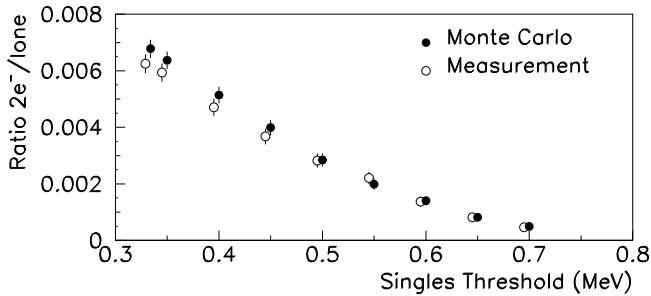}}
\caption{The ratio of 2e$^{-}$ M\"{o}ller events to lone electron 
events as a function of singles threshold.  
Polar angles ($\vartheta$) are restricted to $|\cos \vartheta| < 0.8$.
The measured data are shifted
by -0.01~MeV for clarity.
}
\label{sr90ratio2}
\end{figure}

Measured 2e$^{-}$ events, generated by M\"{o}ller scattering in the 
source plane, can also be compared with the simulated results.
The ratios of scanned 2e$^{-}$ events to lone electron events,
both taken from measured $^{90}$Sr decays,
were calculated at nine singles thresholds.
As seen in Fig.~\ref{sr90ratio2}, the measurements are in agreement with the
Monte Carlo.
Also, from this figure, a systematic 
uncertainty of 10~\% can be estimated and is applied to all Monte Carlo 
results involving electrons. 

\section{TESTS OF DETECTOR RESPONSE MODEL}

The response model of the previous section can be used to calculate 
several specific detection efficiencies which can be tested by direct 
measurement, as summarized in Table~\ref{efncycomp}.

\subsection{Single-Electron Efficiency, $\varepsilon_{e}$}
\label{seeff}

\subsubsection{$\varepsilon_{e}$ from $^{207}$Bi Internal Conversion}
\label{seeffbitos}

The calibrated $^{207}$Bi source, at the center of the $^{150}$Nd 
isotope deposit, has an activity of $140.0 \pm 4.0$ decays per hour.  Given the
conversion rate of the 1064~keV $\gamma$-ray, a 0.6\%
loss due to electron tracks spoiled by X-rays and Auger electrons,
and a further 2.2\% loss due to 
an accompanying second conversion electron, 
$13.4 \pm 0.4$ single K,L,M... electrons from the 1064~keV transition 
can be expected every hour. 
The same calculation for the 
570~keV $\gamma$-ray gives 
$2.98 \pm 0.09$ single electrons per hour.
During
6286.6 hours of livetime, 39\,724 electrons were observed in the
energy range expected from conversion of the 1064~keV $\gamma$-ray,
and 8907 
electrons for the 570~keV transition.
Thus an overall efficiency of $\varepsilon_{e}$=$(47.2\pm1.2)$\% 
can be calculated by combining the number of expected electrons from the two 
lines and the corresponding number of reconstructed tracks.
An independent measurement was made with the much hotter 6~nCi 
$^{207}$Bi source described in Sec.~\ref{ic}.
The activity of 
this uncalibrated $^{207}$Bi source was determined from the TPC 
trigger rate, which was dominated (98\%) by the $^{207}$Bi.  Corrections 
were included for X-rays, Auger electrons, etc.
This method yielded $\varepsilon_{e}$ = 
$(48.6\pm0.8)$\% for the 1~MeV line.  

\subsubsection{$\varepsilon_{e}$ from $^{214}$Bi $\beta$-decay}
\label{eb}

\begin{figure}
\vskip -0.5 cm
\mbox{\epsfxsize=8.6 cm\epsffile{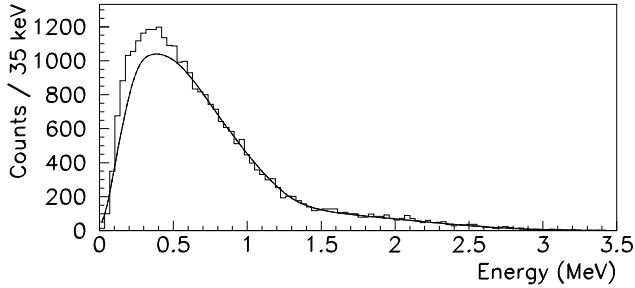}}
\caption{The TPC-measured $^{214}$Bi $\beta$-decay energy spectrum
following $^{222}$Rn injection. 
Solid curve is the GEANT Monte Carlo prediction normalized 
above 0.5~MeV.  The difference arises largely from contamination of 
the event sample by $^{214}$Pb.}
\label{bi214_lone}
\end{figure}

A TPC measurement of the $^{214}$Bi $\beta$ spectrum (Fig.~\ref{bi214_lone})
provides another efficiency check.
Approximately 1~nCi of $^{222}$Rn was 
injected into the TPC to produce a deposit of daughter products, 
including $^{214}$Bi, on the source plane.
Since each $^{214}$Bi $\beta$ decay is followed by the $\alpha$ 
decay of 164 $\mu$s $^{214}$Po, the requirement of a spatially correlated 
$\alpha$ particle in the millisecond following the trigger was 
used to
select triggering $^{214}$Bi events, without regard to the quality of 
the $\beta$ tracks.  
From the resulting 46\,151 event sample,
the single electron analysis software (Sec.~\ref{analysis}) 
reconstructed 9561 events 
originating on the source plane with energies above 
500~keV.  The energy cut was imposed to avoid complications 
related to the threshold of the detector and to a 5.5\% contamination of the 
sample by $^{214}$Pb events.  A correction factor of 1.030 must be 
applied to the 9561 events to account for $\beta$ tracks 
spoiled by overlapping $\alpha$ tracks that satisfy the $\alpha$ 
selection criteria, but have short delays.

Only 51.9\% of the true $^{214}$Bi $\beta$ spectrum survives a 
500~keV cut.  However, the primary sample 
was already depleted in low energy events, since a Monte Carlo 
calculation predicts that $(8.1\pm 0.8)$\% of the electrons fail 
to escape the source and cause a trigger.  
An additional $28.2\pm 2.3$\% correction eliminates contamination of the 
sample 
by $\beta$ tracks originating on the TPC walls rather than the source plane.  
The resulting single-electron efficiency $\varepsilon_{e}=$
\[\frac{9561\times 1.030\times [1-(.081\pm .008)]\times 100}
{46151\times .519\times [1-(.282\pm .023)]}=(52.6\pm 1.6)\%. \]

\subsection{Two-Electron Efficiency, $\varepsilon_{ee}$}

\subsubsection{$\varepsilon_{ee}$ from the $^{207}$Bi decay scheme}
\label{bitosebb}

The small $^{207}$Bi deposit located in the center of each source
assembly can also be used to determine the 2e$^{-}$ efficiency.
Occasionally, two transitions in the cascade
will each yield a conversion electron, producing a 2e$^{-}$ event.
The product of the  conversion rates for the 0.5 and 1~MeV lines gives the
rate of IC-IC pairs as $2.914\!\times\!10^{-3}$ per 1~MeV
$\gamma$-ray.  This implies that an IC-IC cascade should occur at a
rate of $2.157\!\times\!10^{-3}$ per $^{207}$Bi decay. The activity of
the drop on the $^{150}$Nd source is $1.05 \pm 0.03$ pCi, which should
give $0.3017 \pm 0.009$ IC-IC pairs per hour.

\begin{figure}
\mbox{\epsfxsize=8.6 cm\epsffile{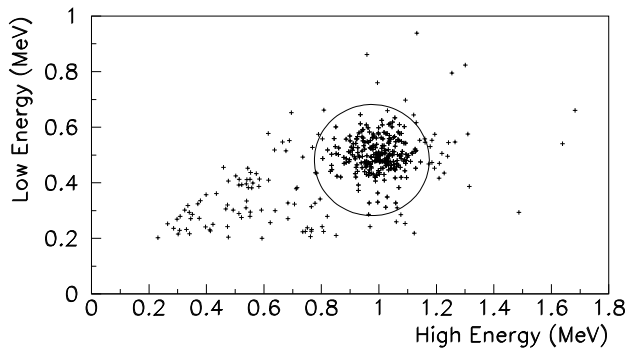}}
\caption{Scatter plot of electron energies 
for events in the $^{150}$Nd 2e$^{-}$ data set which
emerged near the $^{207}$Bi deposit.  
The circle is centered on
481.7~keV and 975.6~keV (the $K$ conversion energies).
M\"{o}ller and Compton scattering contribute events outside the 
circle.}
\label{ndbi207}
\end{figure}

The 2e$^{-}$ stripper used to extract $\beta\beta$-decay events
also saves $^{207}$Bi IC-IC decays which are easily 
tagged by their location at the center the source plane.
In the scatter plot of Fig.~\ref{ndbi207}, most of the points are IC-IC 
events, as is
evident from the clustering near the 481.7~keV and 975.6~keV
conversion energies.  These events are 
selected by a circular cut
which includes 99\% of the
IC-IC signal.  Points outside the circle~\cite{erice} represent M\"{o}ller
scattering of single conversion electrons, or single conversion
electrons accompanied by a Compton electron from a cascade gamma ray.

The selection method described above identified 253 IC-IC events
in a livetime of 6176.6 hours, 
giving an observed rate of $0.041\pm 0.003$ events per hour.  When
compared to the estimated activity of the $^{207}$Bi deposit, this
rate implies an efficiency of $\varepsilon_{ee}=(13.6\pm 0.9)$\%.  

\subsubsection{$\varepsilon_{ee}$ from the $^{214}$Bi decay scheme}

\begin{figure}
\vskip -0.5 cm
\mbox{\epsfxsize=8.6 cm\epsffile{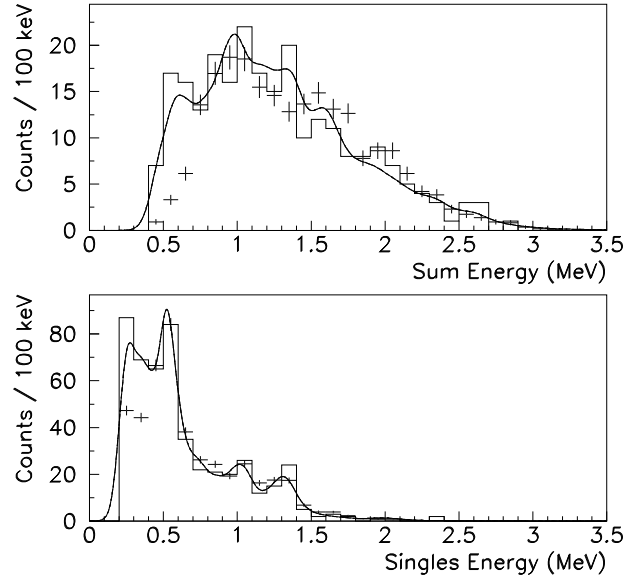}}
\caption{Measured $^{214}$Bi 2e$^{-}$ spectra,
satisfying a 0.2~MeV singles threshold, following $^{222}$Rn 
injection.  Gaussian ideogram (solid curve) is a better
representation of the expected signal than is the histogram. 
Some $^{214}$Pb contamination is present, but is not included in 
the GEANT Monte Carlo predictions 
(crosses).}
\label{bibkgrnd}
\end{figure}

The $^{222}$Rn injection described in Sec.~\ref{eb} yielded 
a sample of 46\,151 $^{214}$Bi decays
from which the offline analysis reconstructed a total of 
189 2e$^{-}$ events (Fig.~\ref{bibkgrnd}).
A single-electron threshold of 400~keV then was imposed on the data
to escape the influence of $^{214}$Pb contamination.
The 64 above-threshold events 
were individually examined, and 
59 good 2e$^{-}$ events were
identified.  

These 59 events included contributions from 
$\beta$ + IC, $\beta$ + M\"{o}ller, $\beta$ 
+ Compton, etc. (See Table~\ref{twoemc}).  Unlike the situation in 
Fig.~\ref{ndbi207}, however, the events involving IC are not easily isolated 
and counted.  Therefore, a 
comparison of the $\beta$ + IC subset to the number of $\beta$ + IC 
events expected from 
the decay scheme is not a viable method of calculating the 
efficiency.  Instead, one must take the quotient of 
all 59 measured 2e$^{-}$ events and the total 
of all types of 2e$^{-}$ events produced in the source.  An estimate 
of the latter requires the very response model that we 
are trying to test by this measurement.  To avoid this circular 
argument, the 
$^{214}$Bi 2e$^{-}$ efficiency is rejected as a useful test. 

Nevertheless, one can still carry out a meaningful check of the model by 
directly comparing the 59 measured 2e$^{-}$ events with the 
number that the model predicts would be measured from 46\,151 $^{214}$Bi 
decays.  The overall conversion rate was calculated by incorporating 
the $\beta$, $\gamma$, and IC characteristics of 
the complicated $^{214}$Bi decay scheme 
in a stand-alone Monte Carlo program.
This simulation considered $>$99\% of the allowed decay channels, and
estimated the conversion rate to be 0.0194 single conversion electron
per $^{214}$Bi $\beta$-decay (Table~\ref{bicprop}).  Feeding this 
information into the response model, with the usual corrections for 
wall events and spoiled tracks described in Sec.~\ref{eb}, 
resulted in a prediction of $52.9\pm 3.4$ 2e$^{-}$ 
events from the 46\,151 decays.  The favorable comparison to the 59 measured 
events is included in Table~\ref{efncycomp}.

\begin{table}
\caption{TPC-measured efficiencies and two-electron events 
as tests of the Monte Carlo model.}
\begin{tabular}{lcccc}
 &  \multicolumn{2}{c}{Single electron} &  \multicolumn{2}{c}{Two-electron} \\
 &  \multicolumn{2}{c}{efficiency $\varepsilon_{e}$ (\%)} &  
\multicolumn{2}{c}{efficiency $\varepsilon_{ee}$ (\%)} \\ \cline{2-3} 
\cline{4-5}
 &Measured &Monte Carlo &Measured &Monte Carlo \\
\tableline  \rule{0mm}{2.75ex} 
$^{207}$Bi & &$53.0 \pm 0.1$  & & $13.9\pm 0.1$  \\
~~1~pCi   & $47.2 \pm 1.2$    &   & $13.6 \pm 0.9$  & \\
~~6~nCi   & $48.6 \pm 0.8$    &   &\multicolumn{2}{c}{Two-electron events}  
\\ \cline{4-5}
$^{214}$Bi     & $52.6 \pm 1.6$    & $54.6 \pm 0.2$   & $59 \pm 8$
& $52.9\pm 3.4$ \\
\end{tabular}
\label{efncycomp}
\end{table}

\subsection{The Model Testing Process}

We test the reliability of the Monte Carlo for 
efficiency and 2e$^{-}$ event prediction in those specific cases 
that have been measured directly, namely $^{207}$Bi and $^{214}$Bi as 
described above.
The angular distributions and energies of 
$\beta\beta_{2\nu}$ are replaced in the Monte Carlo with those for
the appropriate isotope.  For example, the opening angle distribution 
for the IC-IC events of $^{207}$Bi~\cite{kleinheinz},
\[\rm{d}\omega/\rm{d}(\cos\vartheta) \sim 1+0.271\cos^{2}\vartheta \]
and the energies of the two conversion electrons, when inserted in the Monte 
Carlo, give $\varepsilon_{ee}$ for $^{207}$Bi as $13.9\pm 0.1$, vs.~a 
measured value of $13.6\pm 0.9$.  The Monte Carlo and measured values 
for $\varepsilon_{e}$ and $\varepsilon_{ee}$ are summarized in 
Table~\ref{efncycomp}.  These results are consistent with the  10\% 
systematic uncertainty assigned to the Monte Carlo in Sec.~\ref{response}.

\section{THE $\beta\beta_{2\nu}$ EFFICIENCY, $\varepsilon_{\beta\beta}$}

\begin{figure}
\vskip -0.5 cm
\mbox{\epsfxsize=8.6 cm\epsffile{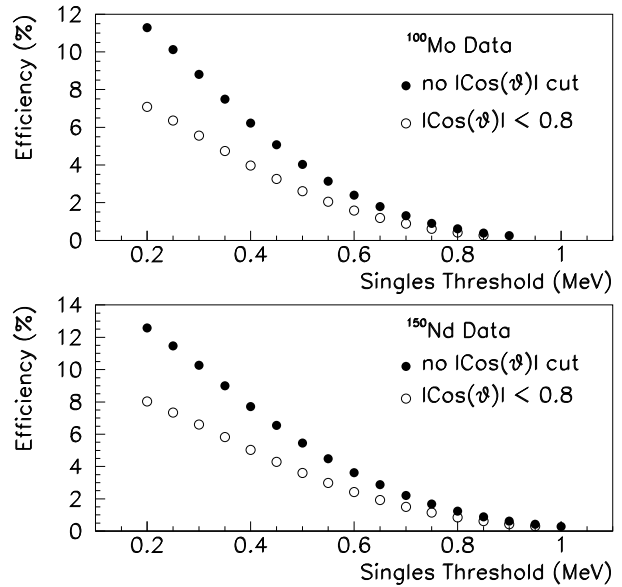}}
\caption{
The efficiency $\varepsilon_{\beta\beta}$ as a function of singles 
threshold.  The statistical uncertainties are negligible compared with the
10~\% systematic uncertainty (not shown).
}
\label{efffig}
\end{figure}

Determination of the $^{100}$Mo and $^{150}$Nd half-lives from 
the observed rates requires knowledge of the $\beta\beta_{2\nu}$ 
efficiency, $\varepsilon_{\beta\beta}$.  Since there is no direct way 
to measure $\varepsilon_{\beta\beta}$, it must be estimated by Monte 
Carlo.  The code employs the detector response model (Sec.~\ref{response}) 
and utilizes the opening angle and energy 
distributions expected for $\beta\beta_{2\nu}$ in the Primakoff-Rosen 
approximation~\cite{primrose}.  The Monte-Carlo-generated efficiencies, as a function 
of singles threshold, are shown in Fig.~\ref{efffig}.

\section{Backgrounds}
\label{backgrounds}

\begin{table*}
\caption{Breakdown of 2e$^{-}$ events, from $10^6\; \beta$-decays,
satisfying a 0.25~MeV singles threshold.  These were generated by the 
GEANT model of intrinsic effects using the simulated decay schemes.
The columns denote combinations of 2e$^{-}$ events
consisting of the beta ($\beta$), Internal Conversion (IC),
M\"{oller} (M), Compton (C) and Photoelectric (P) electrons.
$\star=\beta$ or IC or M or C.  
Combinations with fewer than 10 events are not shown but are included in
``All''. In columns lacking a $\beta$, the $\beta$ did not escape the 
source.}
\begin{tabular}{crrrrrrrrrrrr}
& \multicolumn{6}{c}{$^{100}$Mo} & \multicolumn{6}{c}{$^{150}$Nd} \\ 
\cline{2-7} \cline{8-13}
Isotope & $\beta+\rm{IC}$ & $\beta+\rm{M}$ & $\beta+\rm{C}$ & 
$\rm{IC}+\rm{IC}$ & P + $\star$ & All 
        & $\beta+\rm{IC}$ & $\beta+\rm{M}$ & $\beta+\rm{C}$ & 
$\rm{IC}+\rm{IC}$ & P + $\star$ & All \\
\tableline  \rule{0mm}{2.75ex}
$^{234}$Pa$^{\rm m}$ 	
	&  548 & 260 &   1 & 0  &   0 & 811 
	&  546 & 255 &   1 & 0  &   1 & 804 \\ 
$^{214}$Pb 
	& 2568 &   2 &   0 &  0 &  24 & 2594
	& 2693 &   3 &   0 &  0 &  38 & 2734 \\
$^{214}$Bi   		
	& 2163 & 206 & 167 & 14 &  21 & 2577 
	& 2107 & 164 & 151 & 14 &  36 & 2480 \\
$^{228}$Ac 		
	&  972 & 73 &   50 & 14 &  21 & 1135 
	&  995 & 72 &   59 & 18 &  46 & 1195 \\
$^{212}$Pb  	
	&   10 &  0 &    0 &  0 &   0 &   10 
	&    6 &  0 &    0 &  0 &   0 &    6 \\
$^{212}$Bi  		
	&  234 & 251 &   9 &  6 &   1 &  503
	&  241 & 231 &  17 &  4 &   3 &  497 \\
$^{208}$Tl 		
	& 4501 & 143 & 204 & 37 &  28 & 4924
	& 4527 & 113 & 197 & 44 &  78 & 4966 \\
$^{60}$Co 		
	&    0 &   0 &   6 &  0 &   0 &    7
	&    0 &   0 &   6 &  0 &   3 &   10
\end{tabular}
\label{twoemc}
\end{table*}

Candidates for 
$\beta\beta$-decay are characterized 
by two negative electrons 
with a common point of origin on the $\beta\beta$ source.  
Although this signature 
rejects the vast majority of events from radioactive 
impurities, there are 
certain non-$\beta\beta$ processes that mimic the 2e$^{-}$
topology.  These include M\"{o}ller scattering of single 
electrons born in the source, and 
$\beta$-$\gamma$ cascades in which a $\gamma$-ray Compton scatters, 
produces a photoelectron, or internally converts. (See 
Table~\ref{twoemc}.)
Pair production ($e^{+}e^{-}$) is not a
background since the helical track of the positron would be
rejected by a pitch of the wrong sign.

Dominant among these mechanisms is $\beta$ decay 
accompanied by an internal conversion ($\beta$+IC), a relatively 
common phenomenon in 
the $^{238}$U and $^{232}$Th decay chains.
These families always are 
present at some level in the $\beta\beta$ 
isotope stock.  In addition, daughters of gaseous 
$^{220}$Rn or $^{222}$Rn can migrate to the $\beta\beta$ source 
surface from Ra impurities elsewhere in the chamber or in
the gas-handling system.  

The decay chains supply 
the major radioactivity feeding the other background processes 
as well.
Fortunately, each of these chains contains one or more isotopes
with unique and identifiable decay characteristics, so
good estimates of total uranium and thorium contamination can be
made from the observed activity.

\subsection{The Uranium Series}
\label{useries}

The uranium chain includes several isotopes with high enough 
Q$_{\beta}$ values 
to be of concern.  A summary 
of their properties is included in Table~\ref{bicprop}.

\begin{table}
\caption{Properties of isotopes in the U, Th decay chains 
capable of producing high-energy $\beta$+IC events.  ``Total'' 
includes events with one or more conversion electrons, and 
``Single'' indicates one conversion electron only .
These are the raw conversion rates derived from Monte Carlo simulations 
of the decay schemes, without reference to the TPC.}  
\begin{tabular}{cddd}
 & $Q$-Value & \multicolumn{2}{c}{IC per $\beta$-decay (\%)} \\ \cline{3-4}
Isotope  & (MeV)   & \hspace*{1ex}Total\hspace*{1ex} & 
                    \hspace*{1ex}Single\hspace*{1ex} \\
\tableline  \rule{0mm}{2.75ex}
$^{234}$Pa$^{\rm m}$ 	& 2.29  & 1.27  & 1.18 \\
$^{214}$Pb   	    	& 1.02  & 40.6  & 35.1 \\
$^{214}$Bi   		& 3.27  & 1.95  & 1.94 \\
$^{228}$Ac 		& 2.14  & 80.3  & 67.5 \\
$^{212}$Pb  		& 0.573 & 41.5  & 40.2 \\
$^{212}$Bi  		& 2.25  & 0.343 & 0.339 \\
$^{208}$Tl 		& 4.99  & 7.07  & 6.91  \\
\end{tabular}
\label{bicprop}
\end{table}

Activity below $^{230}$Th is quantified 
by the characteristic 164 $\mu$s $\beta$-$\alpha$ sequence from 
$^{214}$Bi and $^{214}$Po (``BiPo'' events).
The TPC's 1~ms trigger duration includes
99\% of the $^{214}$Po decays.  Alpha tracks that extend into the 
TPC gas are recorded and stored along with the $\beta$ tracks
responsible for the triggers.

The $^{100}$Mo source 
showed a low level of $^{214}$Bi events distributed 
over the entire source plane, irrespective of the region where the $^{100}$Mo 
was located.  This uniformity suggests that very little of the 
activity in the $^{100}$Mo runs arose from contamination 
of the isotope material.
Increased TPC gas flow rates were accompanied by fewer 
observed $^{214}$Bi decays.  The flow dependence 
implied that the primary source of $^{214}$Bi in the $^{100}$Mo 
data set was $^{222}$Rn in the chamber gas.  
Data collected during the 
$^{150}$Nd runs indicate that the Nd$_{2}$O$_{3}$ itself harbored 
additional $^{214}$Bi activity 
roughly equal to that from $^{222}$Rn contamination of 
the chamber gas.  

Quantifying the $^{214}$Bi contamination, however, 
does not constrain $^{234}$Pa$^{\rm m}$.  A breach of secular 
equilibrium could allow a 
$^{234}$Pa$^{\rm m}$ signal without any $^{214}$Bi.  Since the Mo 
refinement and enrichment processes  could throw these isotopes 
out of equilibrium, the possibility of $^{234}$Pa$^{\rm m}$ background 
must be considered.

\subsubsection{$^{214}$Bi Background}
\label{bitfbckgrnd}

$^{214}$Bi with its
3.27~MeV $Q$-value, is potentially the single
most dangerous background in $\beta\beta$-decay experiments with
isotopes like $^{100}$Mo ($Q\!=\!3.03$~MeV) and $^{150}$Nd
($Q\!=\!3.37$~MeV).  The $^{214}$Bi 2e$^{-}$ sum-energy spectrum 
collected during the {$^{222}$Rn}-injection runs 
(Fig.~\ref{bibkgrnd})
is not unlike the expected $\beta\beta_{2\nu}$
signal.  
The $\alpha$ particles and shake-off electrons from decay of the $^{214}$Po 
daughter are exploited to tag 
$^{214}$Bi events with good efficiency (Sec.~\ref{pap}).
The number of tagged events 
together with the 
known tagging efficiency P$_{\alpha}(^{214}$Po) give an accurate measure 
of the number of 
untagged $^{214}$Bi events remaining in the data.
\begin{eqnarray*}
{\rm N}_{\rm untagged} = {\rm N}_{\rm tagged}
(1+\xi)\frac{1-\eta {\rm P}_{\alpha}(^{214}
{\rm Po)}}{\rm{P}_{\alpha}(^{214}{\rm Po})} 
\end{eqnarray*}
where $\xi=0.081$ and $\eta=0.987$ are corrections for corruption of 
Bi events by Po $\alpha$'s in the first 15 $\mu$s, or for Po decays
beyond the 1024 $\mu$s window.
We found that we could also tag 
$^{214}$Bi events by one or more of the three 
{\em previous} links in the uranium chain, with about 50\% efficiency in the 
more difficult $^{150}$Nd case.  The number of $^{214}$Bi events 
remaining after the primary $^{214}$Po tag, however,  was already so small 
(Table~\ref{bgsum}) that this more tedious procedure was not needed.
Another 
tool for reducing the $^{214}$Bi contribution 
is a cut on the strong 1.3~MeV conversion line in the singles 
spectrum, but that technique also was not needed here.

\begin{table}
\caption{Summary of 2e$^{-}$ $\beta\beta$-decay background events. 
$^{228}$Ac, $^{212}$Pb and $^{212}$Bi contamination are estimated from 
$^{208}$Tl activity (see Sec.~\ref{thseries}).
Also listed are M\"{o}ller
scattering contributions as estimated from lone-electron energy spectra.
``Cuts'' include a 250~keV singles threshold and 4-wire radius 
of exclusion about the $^{207}$Bi spot at the center.}
\begin{tabular}{ccccd@{${}\pm{}$}r@{}l}
&Back- & \multicolumn{5}{c}{Number of Events} \\ \cline{3-7}
$\beta\beta$-decay  & ground & \multicolumn{2}{c}{Tagged} 
& \multicolumn{3}{c}{Remain} \\ \cline{3-4}
Isotope &   Type    & Found  & Passed Cuts & \multicolumn{3}{c}{in Data} \\
\tableline   \rule{0mm}{2.75ex}
$^{100}$Mo 
     & $^{214}$Pb  & 24 &  4 & 0.16 & 2 & .09 	\\
     & $^{214}$Bi  &  9 &  4 & 0.12 & 0 & .11 	\\
     & $^{228}$Ac  & -- & -- & 8.8  & 3 & .5	
	\tablenotemark[1]			\\
     & $^{212}$Pb  & -- & -- & 0.08 & 0 & .03 	
	\tablenotemark[1]			\\
     & $^{212}$Bi  & -- & -- & 0.6  & 0 & .2 	
	\tablenotemark[1]			\\
     & $^{208}$Tl  & 14 &  8 & 5.8  & 2 & .6   	\\
     & M\"{o}ller  & -- & -- & 20.5 & 3 & .6  		
	\tablenotemark[1]\tablenotemark[2]    	\\
\hline \rule{0mm}{2.75ex}
$^{150}$Nd 
     & $^{214}$Pb  & 76 & 25 & 33.3 & 10 & .2  	\\
     & $^{214}$Bi  & 61 & 44 &  9.0 &  4 & .7   \\
     & $^{228}$Ac  & -- & -- & 11.9 &  4 & .4
	\tablenotemark[1]			\\
     & $^{212}$Pb  & -- & -- & 0.06 &  0 & .02 
	\tablenotemark[1]			\\
     & $^{212}$Bi  & -- & -- &  0.8 &  0 & .3 
	\tablenotemark[1]			\\
     & $^{208}$Tl  & 25 & 10 &  7.9 &  3 & .3	\\
     & M\"{o}ller  & -- & -- &  9.2 &  3 & .1  	
	\tablenotemark[1]\tablenotemark[2]    	\\
\end{tabular}
\tablenotetext[1]{Monte Carlo systematic uncertainty included.}
\tablenotetext[2]{Excludes $\beta+\rm{M}$\"{o}ller events already 
included for the above isotopes.}
\label{bgsum}
\end{table}

\subsubsection{$^{214}$Pb Background}
\label{pbtfbckgrnd}

Although it has only a 1.0~MeV $Q$-value, $^{214}$Pb  has a very high IC 
probability which makes it a major source of low-energy background.  
The $^{214}$Pb $\beta$+IC singles- and sum-energy spectra 
were measured with the same $^{222}$Rn injection data used to measure 
the $^{214}$Bi spectrum in Sec.~\ref{eb}.  In this case,
however, only manually scanned events with no accompanying 
$\alpha$-particle or shake-off electron were considered.  This sample 
contains $<2$\% contamination of $^{214}$Bi events.

The $^{214}$Pb/$^{214}$Bi ratio of 2e$^{-}$ events from Table~\ref{twoemc}
was used to determine the number of 
$^{214}$Pb events in the $\beta\beta$ data set from the number of tagged plus 
untagged $^{214}$Bi events discussed in the previous section.
(As a check, this ratio, measured from the 
$^{222}$Rn injection data at a 0.200~MeV singles threshold, yielded 
$2.51\pm 0.30$, in agreement with the corresponding Monte Carlo value of 
$2.82\pm0.07\pm0.40$.)

\begin{figure}
\mbox{\epsfxsize=8.6 cm\epsffile{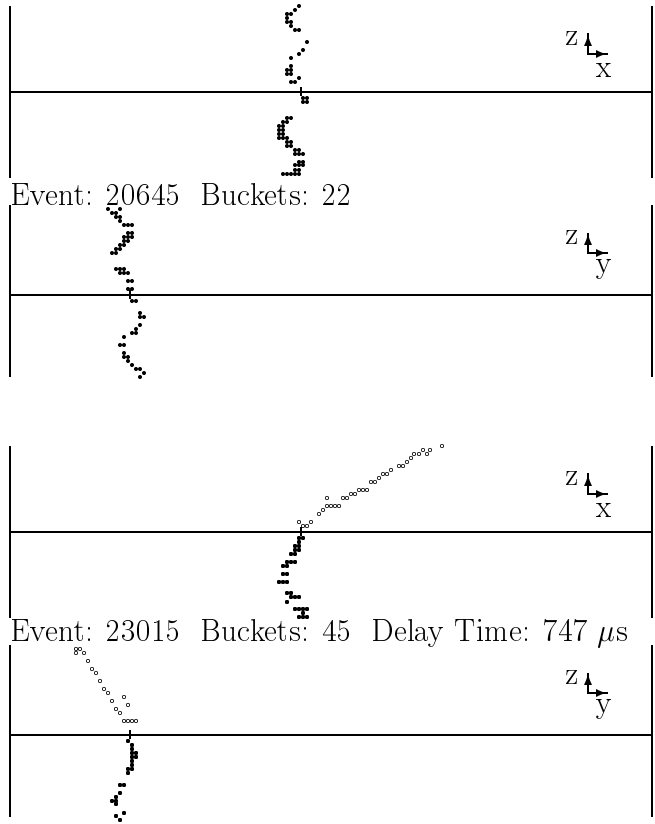}}
\vskip 4mm
\caption{An example of $^{214}$Pb $\beta$+IC
event identification.  Top event is a low-energy $\beta\beta$-decay
candidate.  Bottom event is a ``BiPo'' sequence 
emerging from the same point on the source 82 minutes later,
indicating that the candidate is probably a $^{214}$Pb $\beta$+IC
event.  Tick marks near the vertex are at the same position in each figure. 
Open circles are the $^{214}$Po $\alpha$ track, 747 $\mu$s delayed.}
\label{pbbicex}
\end{figure}

It is possible to identify individually and remove many of the $^{214}$Pb 
events in the $\beta\beta$-decay  data set by tagging them with the
characteristic ``BiPo''
signature that follows with a mean delay of 28.4 minutes.  
$^{214}$Bi decays occur at a rate of only a 
few per hour over the entire source area, so the chance of an 
accidental spatial and temporal coincidence between a $\beta\beta$ 
event and a $^{214}$Bi decay is very small.  If a 1 cm$^{2}$ area of 
coincidence and a two-hour search window are assumed, the chance is 
only about 1 in 500 that an event might be wrongly rejected.  The 
search has been carried out on all of the $\beta\beta$ candidates from 
both sources.  For each candidate, the software searches the raw data 
file for any nearby event which shows a delayed hit pattern indicative 
of an $\alpha$-particle, and occurs within the two hour (six half-life)
time period following the candidate.  After human confirmation of a valid 
``BiPo'' tag (Fig.~\ref{pbbicex}) the $^{214}$Pb event is removed from the 
data.  Results are shown in Table~\ref{bgsum}.

\subsubsection{$^{234}$Pa$^{\rm m}$ Background}

Likely disequilibrium with the rest
of the uranium chain forces us to evaluate
the level of $^{234}$Pa$^{\rm m}$ contamination by searching for its decays
directly.  This isotope has only one strong conversion line at 
695~keV, so any 2e$^{-}$ contribution
should populate the singles-energy spectrum at that energy.  There is
no obvious excess observed at 695~keV in either the $^{100}$Mo or $^{150}$Nd 
2e$^{-}$ singles-energy spectrum, so any $^{234}$Pa$^{\rm m}$
contribution is small.
                                            
The $^{234}$Pa$^{\rm m}$ events that do occur 
must be preceded by the low-energy
$^{234}$Th $\beta$-particle within several 1.2~min half-lives. 
Background identification techniques used to tag $^{208}$Tl 2e$^{-}$ 
events will also find this decay sequence
with high efficiency (although it would probably be labeled as a
$^{208}$Tl event).  Therefore,
the background contribution from $^{234}$Pa$^{\rm m}$ is assumed to be
negligible.
                                                                              
\subsection{The Thorium Series}
\label{thseries}

\begin{figure}
\vskip -0.5 cm
\mbox{\epsfxsize=8.6 cm\epsffile{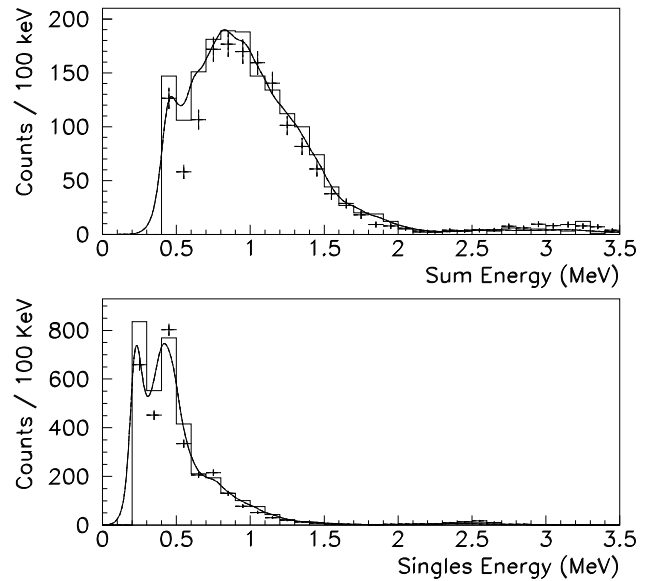}}
\caption{Measured $^{212}$Pb + $^{212}$Bi + $^{208}$Tl equilibrium 
2e$^{-}$ spectrum, satisfying a 0.2~MeV singles threshold, 
following $^{220}$Rn injection.
Solid curve is the Gaussian ideogram.  The corresponding GEANT Monte Carlo 
prediction (crosses) includes the branching and loss of $^{212}$Bi events 
spoiled by the $^{212}$Po $\alpha$.}
\label{tlbkgrnd}
\end{figure}

The potential of thorium daughters for serious $\beta\beta$ background 
is evident in Fig.~\ref{tlbkgrnd} which shows measured 
singles- and sum-energy spectra for equilibrium $^{212}$Pb + $^{212}$Bi + 
$^{208}$Tl, obtained by injecting $^{220}$Rn gas into the center of the TPC.
Also shown are corresponding spectra generated by a Monte Carlo.
Noticeable in the sum-energy spectrum is a significant population in 
the 3~MeV region where one looks for 
$\beta\beta_{0\nu}$-decay in $^{100}$Mo and $^{150}$Nd.  

The presence of thorium chain contaminants can be recognized from 
observation of the 0.3 $\mu$s, mass-212 ``BiPo'' sequence.
Within the temporal resolution of the TPC, both the $\beta$- and
$\alpha$-decays will appear to occur simultaneously, and will both be
recorded in the trigger hits of the event.  This 
signature can be imitated by 
$\approx$1\% of the $^{214}$Bi decays in which the 164~$\mu$s
$\alpha$-decay of $^{214}$Po occurs in the first few  microseconds.
A small correction for the mass-214 
effect can be applied once the uranium chain activity is
estimated, and does not interfere with using $^{212}$Bi as a thorium
tracer.

Unlike the uranium chain
contaminants, thorium activity is not 
distributed over the $\beta\beta$ source by Rn in the
gas supply.  Gas is introduced slowly into the TPC from the top.
As demonstrated by a separate injection of $^{220}$Rn 
into the TPC gas feed line, the charged daughter ions of this short-lived radon 
collect near the point of entry, and subsequent activity is localized 
to the top of the chamber.  
Since this pattern is not observed in normal data runs,
the thorium activity is within the chamber, probably in the 
$\beta\beta$ source itself.

The serious background isotopes of the thorium series are
separated by a maximum half-life of 1.9 years ($^{228}$Th).  Considering
the elapsed time between processing and measurement of our isotopes, it is
safe to say that the thorium chain from $^{228}$Ac to
$^{208}$Tl is in approximate equilibrium.

\subsubsection{$^{208}$Tl Background}

With a $Q$-value of almost 5~MeV and 0.5\% of decays generating two electrons 
passing the standard cuts (Table~\ref{twoemc}), $^{208}$Tl is another 
particularly dangerous background.  Fortunately, the highest-energy 
decays are almost exclusively
from internal conversion of the 2.61~MeV $\gamma$-ray, so
events near 3~MeV can be removed by a cut on
single electrons near the 2.53~MeV conversion energy, without
significantly impacting the rest of the spectrum.

The use of the mass-212 ``BiPo'' count to quantify the activity of the 
Th series is not as straightforward as in the mass-214 case.  The 
effectively instantaneous $^{212}$Po $\alpha$-particle is often lost 
within the $^{212}$Bi $\beta$ track.
It is sometimes
possible, however,  to identify a $\beta\beta$-decay candidate as
a $^{208}$Tl event by a back search for the
$\alpha$-decay which must precede it with a 3.1 minute half-life.  
For all $\beta\beta$-decay candidate events,
the preceeding 20 minutes (six 
half-lives) is searched by software for any particle originating near the 
candidate's vertex.  
Particle identification is left to the 
judgement of the person analyzing the data. 

The efficiency of this search procedure is the probability of
seeing the $^{212}$Bi $\alpha$-decay, $P_{\alpha}(^{212}\rm{Bi})$ 
from Table~\ref{aranges}.  
(This search would be enhanced by the 
conversion electron from the 40~keV level in $^{208}$Tl 
which occurs
with 70\% of the $^{212}$Bi $\alpha$-decays, but because of its low 
energy this electron rarely reaches the TPC gas.)
The corresponding 
$^{208}$Tl count is included in Table~\ref{bgsum}.

\subsubsection{$^{212}$Bi Background}

Although Q$_{\beta}$ is 2.25~MeV, 2e$^{-}$ production per $\beta$ 
decay by $^{212}$Bi 
is only 10\% of that by $^{208}$Tl (Table~\ref{twoemc}).  Branching 
increases the $^{212}$Bi portion to 18\%, but finally it is reduced 
to 4\% by corruption from 
the essentially simultaneous $^{212}$Po $\alpha$-particle, which appears 
in the gas 80\% of the time (Table~\ref{aranges}).  This small background 
contribution of $^{212}$Bi is also included in Table~\ref{bgsum}.

\subsubsection{$^{212}$Pb Background}
\label{pbbckgrnd}

The number of $^{212}$Pb 2e$^{-}$ events can be determined from $^{208}$Tl 
decays in a manner similar to that for $^{212}$Bi above.
The branching fractions and 2e$^{-}$ probabilities predict that 
$^{212}$Pb events should be detected $<0.6\%$ 
as frequently as similar $^{208}$Tl decays.  
These are also included in Table~\ref{bgsum}.

\subsubsection{$^{228}$Ac Background}
\label{acbckgrnd}

The complicated decay scheme of $^{228}$Ac almost always produces an
internal conversion, and will frequently produce {\em
multiple} conversions.
Decays with three or more electrons appearing will not be
accepted as $\beta\beta$-decay events, so this characteristic helps
reduce the probability of $^{228}$Ac contributing to the measured
$\beta\beta$-decay  spectra.

Although the singles-energy Monte Carlo spectrum for $^{228}$Ac 
$\beta$+IC events
has a distinctive peak near 0.85~MeV, there is no hint of such a peak 
in the measured 2e$^{-}$ spectra.

We can estimate the $^{228}$Ac contribution from the $^{208}$Tl activity 
as described in the previous sections.  Assuming that the thorium chain
is in equilibrium, branching fractions and 
2e$^{-}$ probabilities
predict that $^{228}$Ac 2e$^{-}$ events 
should be detected with $\sim 65\%$ the frequency of similar 
$^{208}$Tl decays.  These are also included in Table~\ref{bgsum}.

\subsection{Other Sources of Conversion Backgrounds}

Moe and Lowenthal~\cite{moe80} have investigated the
possibility of $\beta$+IC backgrounds produced by isotopes outside of
the natural decay series.  They considered 
nuclei produced from cosmic ray activity, contaminants from atmospheric 
nuclear weapons test fallout, common
man-made materials such as $^{60}$Co, and 
long-lived non-series isotopes such as $^{40}$K.  In order to be considered a
serious source of $\beta$+IC background, an isotope must have
significant internal conversion probability and $Q$-value, 
a half-life greater than the order of six months
or, alternately, a plausible continuous production
mechanism. (We have never seen a $\beta\beta$ candidate rate decline 
over the duration of the experiment.)  None of the nuclides investigated 
satisfied all of these requirements.


\section{Analysis of the Data}                            
\label{analysis}

\subsection{Off-Line Event Selection}
                            
Each day the data acquisition system recorded about 50\,000 events,
from which a $\beta\beta$ signal of 2 or 3 events had to be 
extracted.  Automated event stripping routines applied a set of empirical
criteria engineered to select events which could be
$\beta\beta$-decay candidates.  A similar set of
routines was designed for identifying lone electrons.

\subsection{Event Scanning and Fitting}

\subsubsection{Lone Electron Events}
\label{lonefits}

Since $^{100}$Mo and $^{150}$Nd are
stable against single $\beta$-decay, all lone electrons spawned in 
the source are due to unwanted radioactivity.  Producers of 
2e$^{-}$ backgrounds generally emit far more copious numbers of 
single electrons.  Studying the lone electrons\footnote{
It was his last wish before his death in 1990, that Prof. D.~Skobeltzin 
of INR Moscow, hoped someone would investigate the anomalous scattering of 
$\beta$ particles observed in his early cloud chamber experiments.  
Suggesting that a He-filled TPC was the ideal detector, Prof. 
G.~Zatsepin relayed this wish to us, and requested that we publish this 
footnote.  Skobeltzin's photographs (eg. Nature {\bf 137}, 234 (1936)) 
showed $\beta$ particles from $^{214}$Bi suffering apparent inelastic 
collisions in which they lost as much as 90\% of their energy.  
As Skobeltzin described it in a book honoring the 60th Anniversary of the
birth of Prof. S.~Vavilov published by the Academy of Sciences of the 
USSR in 1952, the phenomenon occurred in 5-10\% of the $\beta$ tracks 
within 20 cm of the source, and was consistent kinematically with a 
$\approx 10^{-10}$ sec in-flight decay of an unstable particle.  We have 
never observed this effect in the many beta tracks examined in the course 
of our $\beta\beta$ work, and would certainly have seen it at the 5-10\% 
level.  Nevertheless, to be quantitative, we specifically examined the first 
20~cm of 230 high-energy $\beta$ tracks from the decay of $^{214}$Bi. 
We saw no event with a sudden, large loss of energy.  This result rules 
out more than a 1\% effect in our gas mixture, with 90\% confidence.  
Without seeing Prof. Skobeltzin's original photographs, we can only speculate
that the anomaly resulted from nonuniform illumination of his chamber.
}
emitted from
the source plane helps establish limits on the number of associated
2e$^{-}$ background events.  This can be accomplished by fitting the lone 
electron energy spectrum with component background spectra to estimate each 
component's activity.

\begin{figure}
\vskip -0.5 cm
\mbox{\epsfxsize=8.6 cm\epsffile{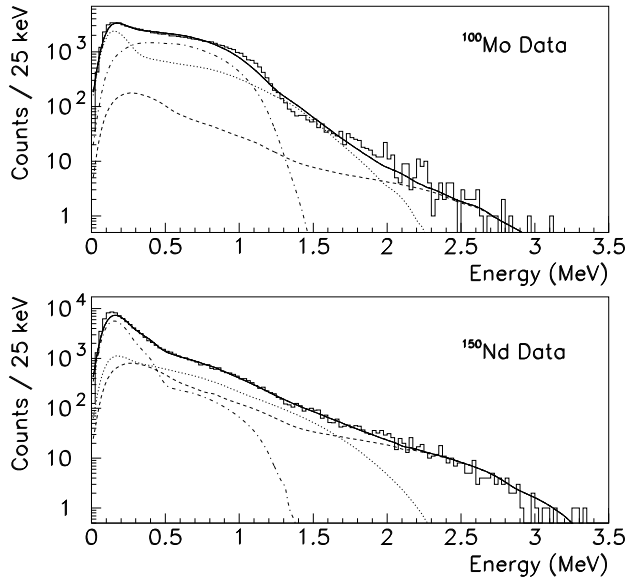}}
\caption{Histograms of the lone electron energy spectra 
from $^{100}$Mo and $^{150}$Nd runs, and their fitted 
(at $\geq$200~keV) background components.
These are $^{238}$U (excluding $^{234}$Pa$^{\rm m}$) (dashed curve), 
$^{232}$Th (dotted curve), the sum of $^{40}$K, $^{60}$Co and 
$^{137}$Cs (dashed-dotted curve), and the total spectrum (solid curve).}
\label{ndlonee}
\end{figure}

Figure~\ref{ndlonee} shows the energy spectra of
lone electrons, whose track vertices are between 10 and 53 wires 
from the $^{207}$Bi source, collected from all but 518 hours of the $^{100}$Mo 
runs and from the entire set of $^{150}$Nd runs.
These spectra were 
fitted with various background lone electron spectra generated by the 
GEANT Monte Carlo using Ref.~\cite{betaspec} for input spectra.
Background component spectra considered were $^{40}$K, $^{60}$Co,
$^{137}$Cs and daughters of the $^{238}$U, $^{232}$Th and $^{235}$U decay 
chains.  In addition, $^{108}$Ag was considered for the 
$^{100}$Mo data while $^{152}$Eu and $^{154}$Eu were included in the fit 
to the $^{150}$Nd data.

The fit to the $^{100}$Mo spectrum was of limited success.
Difficulty reproducing the shoulder near 1.3~MeV is suggestive of a 
contribution from an unidentified contaminant.  We do not 
consider this fit to be valid.  Since the likely heavy elements have 
been included, the unknown contaminant is not expected to be 
a strong IC emitter.  Its dominant contribution 
to 2e${-}$ background should be M\"{o}ller scattering, which is 
accounted for by the lone electron spectrum, independent of any fit 
(Sec.~\ref{bgid}).

The fit of the $^{150}$Nd spectrum, however, yielded a good match to 
the measured data.  The results are shown in 
Table~\ref{loneefitres}.  
Although $^{212}$Bi is in equilibrium with the $^{232}$Th decay chain,  it
was fitted as an independent parameter in order to calculate the $^{212}$Po 
$\alpha$-particle escape probability.  This was determined to be 
$67.5^{+8.2}_{-4.9}$\%, within $1.64\sigma$ of the values listed in
Table~\ref{aranges}.

In addition to the isotopes listed in this table, others were 
considered and rejected. Daughters of $^{235}$U, 
specifically $^{211}$Pb and $^{207}$Tl, were acceptable to the fit but at the 
expense of $^{40}$K.  Ultimately, $^{235}$U background was rejected by 
the fits to 
the 2e$^{-}$ spectra discussed in Sec.~\ref{emlintro}.  Similarly 
$^{152}$Eu was also acceptable in the lone electron fits but not in the 
2e$^{-}$ counterpart. On the other hand, $^{154}$Eu was rejected by the 
lone electron fit.

\renewcommand{\arraystretch}{1.5}
\begin{table}
\caption{Activities corresponding to the fit to the $^{150}$Nd lone 
electron energy 
spectrum of Fig.~\ref{ndlonee}.  Also shown are total 2e$^{-}$ 
events surviving standard cuts, as calculated from the activity, and
from Table~\ref{bgsum} which is considered to be more reliable.}
\begin{tabular}{cr@{}l@{$^{+}_{-}$}lr@{}l@{$^{+}_{-}$}lr@{}l@{${}\pm{}$}r@{}l}
	& \multicolumn{3}{c}{Activity} & 
	\multicolumn{7}{c}{Total 2 e$^-$ events} \\  \cline{5-11}
Isotope	& \multicolumn{3}{c}{($\mu$Bq/g)} & 
	\multicolumn{3}{c}{From Fit\tablenotemark[1]} &
	\multicolumn{4}{c}{From Table~\ref{bgsum}} \\
\tableline   \rule{0mm}{2.75ex}
$^{234}$Pa$^{\rm m}$ 			& $\leq$ 18 & 
			\multicolumn{2}{l}{.7 \tablenotemark[2]}  
		& $\leq$ 8 & \multicolumn{2}{l}{.7 \tablenotemark[2]}	
		& \multicolumn{4}{c}{---} 			\\
$^{214}$Pb 			& 71 & .5 & $^{4.1}_{4.1}$  \tablenotemark[3]
		& 68 & .5 & $^{8.0}_{8.0}$			
		& 58 & .3 & 10 & .2 				\\
$^{214}$Bi 			&  \multicolumn{3}{c}{c}
		& 58 & .3 & $^{6.8}_{6.8}$
		& 53 & .0 & 4 & .7  				\\
$^{210}$Bi 			& 4 & .3 & $^{10.8}_{0.0}$ 
		& 0 & \multicolumn{2}{c}{}				
		& \multicolumn{4}{c}{---} 			\\
$^{228}$Ac 			& 98 & .3 & $^{14.9}_{14.7}$ \tablenotemark[4] 
		& 41 & .1 & $^{7.6}_{7.5}$
		& 11 & .9 & 4 & .4\tablenotemark[1]  	 	\\
$^{212}$Pb 			& \multicolumn{3}{c}{d}
		& 0 & .21 & $^{0.09}_{0.09}$
		& 0 & .06 & 0 & .02\tablenotemark[1]		\\
$^{208}$Tl 			& \multicolumn{3}{c}{d}
		& 61 & .6 & $^{11.2}_{11.1}$
		& 17 & .9 & 3 & .3 				\\
$^{212}$Bi			& 20 & .4 & $^{4.1}_{0.2}$
		& 0 & .7 & $^{0.2}_{0.1}$
		& 0 & .8 & 0 & .3\tablenotemark[1]		\\
$^{40}$K			& 43 & .7 & $^{10.5}_{10.5}$
		& 0 & \multicolumn{2}{c}{}				
		& \multicolumn{4}{c}{---} 			\\
$^{60}$Co			& 613 & & $^{24}_{24}$
		& 2 & .1 & $^{0.7}_{0.7}$				
		& \multicolumn{4}{c}{---} 			\\
$^{137}$Cs			& 239 & & $^{10}_{10}$
		& 0 & \multicolumn{2}{c}{}				
		& \multicolumn{4}{c}{---} 			\\
\end{tabular}
\tablenotetext[1]{Monte Carlo systematic uncertainty included.}
\tablenotetext[2]{Limit at 90 \% confidence level.}
\tablenotetext[3]{\,$^{214}$Pb and $^{214}$Bi were fitted together.}
\tablenotetext[4]{\,$^{228}$Ac, $^{212}$Pb and $^{208}$Tl were fitted 
together.}
\label{loneefitres}
\end{table}

\subsubsection{Two Electron Events}
\label{twoescan}

The offline stripper selects negative electron pairs
emerging from opposite sides of the source plane.   
We did not reconstruct
same-side events because of 
added complexity in event selection and
fitting, as well as in understanding the associated efficiencies.

The stripper attempts to 
reject events which are not $\beta\beta$-decay candidates.  
When it cannot reject an event with certainty, it will
save it.  This system ensures that essentially all recognizable
$\beta\beta$-decay events are accepted, and efficiency of the offline
stripper is not an issue.  Studies conducted during the development of
the stripper showed that it did not reject any events which would not
have been rejected by a human scanner.

The offline stripper saved 2e$^{-}$ ``$\beta\beta$-decay
candidate'' events at rate of about 25--30 per day of running.  These
events were individually studied by a physicist.  Most events
were rejected immediately 
as a multi-tracked
backscattered lone electron, an indecipherable swath of ionization, or
some other two-sided event which was passed by the stripper's policy
of accepting events that it cannot categorize.

Events that have the correct $\beta\beta$-decay topology are fitted
with two independent helices.  The scanning software determines the
energy and opening angle of the event based on the parameters of the
fit in conjunction with the magnet current, drift field voltage and
atmospheric pressure information recorded with the event.  If the fit
is acceptable, the person
scanning the data can elect to save the event.

A 250~keV threshold is imposed on each electron.
Two-electron events from the $^{207}$Bi deposit 
(Sec.~\ref{bitosebb}) are
eliminated by excluding electrons within a 4 wire radius
of the source center.

Occasionally, the fitter will have trouble converging to a good fit
because of small defects in a visually acceptable track.  The scanning
software allows the user to remove these defects.  This is
done only when necessary, and always to the smallest possible
extent.  In the case of gaps in the track, a single added point will
usually guide the fitter to the correct result.  Ionization electrons 
formed where the track enters the 5~mm dead space 
between the TPC wall and the first plane of
wires can arrive at the anode a few $\mu$s later, causing stray hits 
which may be safely deleted.
An electron which scattered abruptly 
in the gas is usually   handled by deleting an entire section
of the track, leaving a single sinusoid for the fitter to work on. 
A status
flag is recorded with the fit parameters indicating those few cases 
where the track was modified.

\subsection{Background Identification}
\label{bgid}

\begin{figure}
\vskip -0.5 cm
\mbox{\epsfxsize=8.6 cm\epsffile{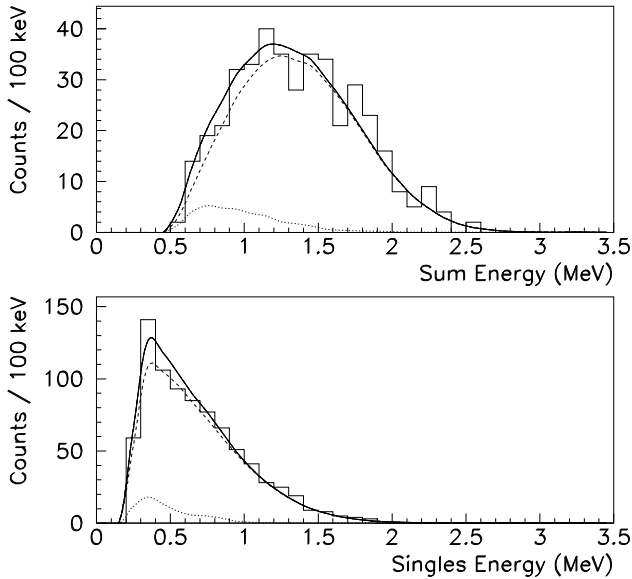}}
\caption{Histograms of the $^{100}$Mo sum- and singles-energy spectra 
for 2e$^{-}$ events with tagged background decays removed. 
Events represented have
single-electron energies greater than 250~keV and originate at
least four wires from the $^{207}$Bi calibration deposit.  
Also displayed are the results of the Extended Maximum Likelihood fit where 
the curves denote total background (dotted), 
$\beta\beta_{2\nu}$ (dashed) and their sum (solid).
}
\label{mofitspc}
\end{figure}

\begin{figure}
\vskip -0.5 cm
\mbox{\epsfxsize=8.6 cm\epsffile{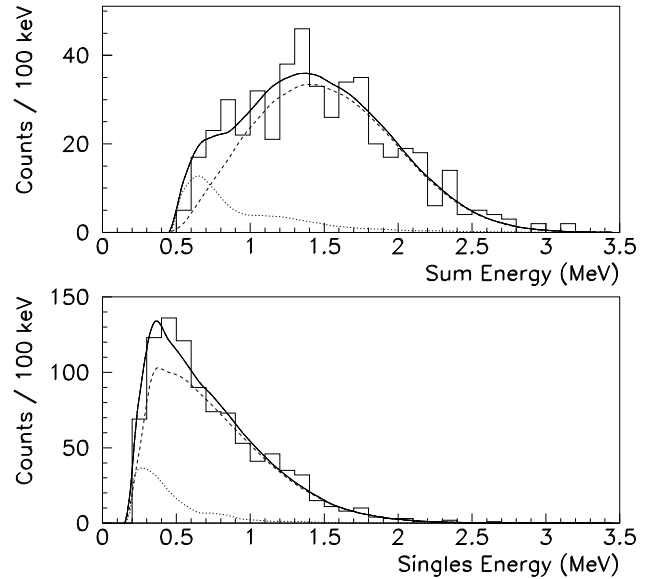}}
\caption{The $^{150}$Nd counterpart of Fig.~\ref{mofitspc}}
\label{ndfitspc}
\end{figure}

The tagging methods described in Sec.~\ref{backgrounds} were used to 
eliminate, event-by-event, much of the U- and Th-series background, 
and estimate the number of background events remaining among the 
$\beta\beta$ candidates.  In the $^{100}$Mo data set 16 background 
events were found, and in the $^{150}$Nd data set 79 events,
which otherwise would have been accepted as $\beta\beta$-decay candidates.
The energy spectra of remaining events 
are shown in Figs.~\ref{mofitspc} and~\ref{ndfitspc}. 
There are 410 remaining 2e$^{-}$ events in the $^{100}$Mo data set 
while the corresponding number for $^{150}$Nd is 476.
Table~\ref{bgsum} shows details of how many background counts were identified
and also summarizes the expected contamination remaining in the data.  
The spectra still include the untagged background events listed 
in the last column of the table. 

The expected background contamination for $^{150}$Nd can also be estimated 
from the activity derived from the lone electron fits.  GEANT 
$\alpha$-particle escape probabilities for $^{212}$Po and $^{214}$Po 
(Table~\ref{aranges}) and 2e$^{-}$ production 
coefficients (Table~\ref{twoemc}) were used in this calculation.
The results, listed in Table~\ref{loneefitres}, are in rough agreement 
with those in Table~\ref{bgsum}, with the exception of the Thorium 
daughters where the fit is less solid due to the interplay of several 
constituents in the same energy region.  These fit results are not 
used in our final background estimates, but serve as a rough 
consistency check of the more precise determinations in Table~\ref{bgsum}.

In addition to U and Th backgrounds, isotopes outside these
series contribute 2e$^{-}$ events through the $\beta \: + $~M\"{o}ller 
process.  M\"{o}ller scattering by the entire lone electron spectrum
can be estimated by Monte Carlo.  Since a subset of these M\"{o}ller 
events has already been accounted for in the contributions of isotopes 
listed in Table \ref{bgsum}, those contributions are subtracted to 
yield the balance labeled ``M\"{o}ller'' in the table.  

The method used to estimate the M\"{o}ller production by the lone 
electron spectrum is as follows.  The true number of lone electrons $N(K)$,
as a function of true energy $K$, can be obtained from the 
measured number $N^{\prime}(K^{\prime})$ where $K^{\prime}$ is the 
measured electron energy.
This is given by 
\begin{equation}
N(K) = \frac{N^{\prime}_{\rm shift}(K)}{g(K)}
\label{truespec}
\end{equation}
where 
\begin{eqnarray*}
N^{\prime}_{\rm shift}(K) & = & \int N^{\prime}(K^{\prime}) \: 
P(K,K^{\prime}) \: dK^{\prime}, \\
g(K) & = & \frac{N^{\rm mc}_{\rm shift}(K)}{N(K)}.
\end{eqnarray*}
$\int P(K,K^{\prime}) \: dK^{\prime} = 1$ 
is a probability distribution function while 
$N^{\rm mc}_{\rm shift}(K)$ is the number of events surviving; 
both were determined from Monte Carlo using the measured 
$^{100}$Mo lone energy spectrum as input.  
The first iteration true spectrum, obtained in this manner,
was input to the Monte Carlo to obtain the lone and 2e$^{-}$ energy 
spectra.  The former was found to be in excellent agreement with the 
measured spectrum; thus further iterations were unnecessary.

The total $\beta \: + $~M\"{o}ller contribution, estimated from the Monte Carlo,
is $21.8\pm2.7\pm2.3$ for $^{100}$Mo and $13.9\pm2.6\pm1.5$ for
$^{150}$Nd.  The $\beta \: + $~M\"{o}ller events, remaining in the 
data and listed in Table~\ref{bgsum}, were obtained by subtracting 
the contribution of each isotope in the table. This contribution was 
determined by scaling the remaining events by the Monte Carlo 
determined fraction of $\beta \: + $~M\"{o}ller to total 2e$^{-}$ 
events.

\subsection{Determination of $\beta\beta$-decay Rate}

The spectra of $\beta\beta$-decay candidates remaining are composed of 
real $\beta\beta$-decay events and a residue of 2e$^{-}$ 
backgrounds.  Good half-life determinations depend on accurate
appraisals of the signals contained in the spectra.  This can be done
in several ways.
            
\subsubsection{Cuts Applied to Observed Spectra}

The majority of 2e$^{-}$ backgrounds occur at low energies, and 
can be greatly reduced by a high singles-energy threshold.
A Monte Carlo simulation can then be used to estimate the
number of events which would survive the cut, allowing us to 
estimate the number of events in the entire spectrum based on those
events over our threshold.

A 500~keV singles-energy cut is a reasonable threshold 
choice, as it eliminates a large portion of the backgrounds while
retaining much of the $\beta\beta$-decay.
Higher energy cuts would remove somewhat more background, but they rapidly
degrade the statistics of the measured $\beta\beta$-decay spectra.
This cut results in estimates for the standard uncut spectrum, of
$420\pm 34$ events in the $^{100}$Mo energy spectrum, and 
$480\pm 33$ counts in the $^{150}$Nd spectrum.  
(This same cut applied to the
$|\cos\vartheta|\!<\!0.8$ spectrum implies 
$407\pm 42$ $^{100}$Mo events, and $423\pm 39$ $^{150}$Nd events.)

\subsubsection{Fits to the Observed Energy Spectra}
\label{emlintro}

In addition to the above method, the data were fitted with
the spectra expected from both the signal and the common backgrounds.  This
method allows full use of all data.

The method employed for the fits was the Extended Maximum Likelihood (EML)
method~\cite{barlow90,Lyons,orear58}.  EML analysis uses one or more
probability distributions which should describe the data in question, each of
which has one or more free parameters describing its shape.  The
parameters are adjusted so that the probability of the data being
derived from that distribution is maximized.

For spectral fits, Monte-Carlo-generated $\beta\beta$ and
background spectra are used as probability distributions, with
the only free parameters being their absolute normalizations.  
Exceptions were $^{208}$Tl, $^{212}$Pb and $^{212}$Bi for which 
the measured $^{220}$Rn injected spectrum was used.
Opening angle information was
not utilized in the fitting procedure (see 
Sec.~\ref{opangacc}).

The likelihood function is given by
\begin{equation}
{\cal L} = \left[\prod_{i} P(x_{i},\vec{a})\right] e^{-{\cal N}}
\label{eml}
\end{equation}
where $P(x_{i},\vec{a})$ is the probability of observing event $x_{i}$, given
the distribution described by the parameters $\vec{a}$, and ${\cal N}$ is
the normalization of the model.  More specifically, the probability
$P(x,\vec{a})$ is the sum of all the probabilities of the normalized spectra
of interest, i.e.\
\begin{eqnarray*}
P(x,\vec{a}) = \sum_{j} a_{j}p_{j}(x),
\end{eqnarray*}
where
\begin{eqnarray*}
\int p_{j}(x)\;dx = 1.
\end{eqnarray*}
The $p_{i}(x)$ may include models
of a variety of energy spectra, i.e.\ $\beta\beta_{2\nu}$-decay sum,
$\beta\beta_{2\nu}$-decay singles, $\beta\beta_{0\nu,\chi}$-decay sum,
$\beta\beta_{0\nu,\chi}$-decay singles, $^{214}$Pb sum, $^{214}$Pb
singles, etc., as well as other distributions such as that
of $\beta\beta$-decay opening angles.  The parameters $\vec{a}$ are
then explicitly the absolute normalizations of the various model
components, so that
\begin{eqnarray*}
{\cal N} = \sum_{i}a_{i}.
\end{eqnarray*}
When both singles- and sum-energy spectra are used in a fit, a single
normalization is used to describe both spectra; i.e., there is a
single parameter used to describe the number of events attributed to
each process.  With this scheme, the sum-energy spectrum
normalization for a given background process would be $a_{j}$, while
the normalization of the associated singles-spectrum would be
$2a_{j}$.

A fitter was developed for maximizing Eq.~(\ref{eml}) based on the
CERN Program Library's Minuit minimization package. The fitter was
designed for maximum flexibility, and allows the user to select
arbitrary combinations of models, apply various cuts to the data,
restrict fits to specified energy ranges, and even apply a random
smearing of the data for studies of systematic effects.

\subsubsection{Inclusion of Background Estimates}
\label{ibe}

One has to choose a reasonable method for applying the various measured and 
calculated background models.  Two extreme
possibilities are either to fix all background model normalizations at
the best estimates while leaving only the $\beta\beta$-decay model
free to fluctuate in the fit, or to leave all models completely
unconstrained.

Neither of these extremes is particularly appealing.  While the
background estimates are based on careful measurements, and
there is no particular reason to doubt them, the
$\beta\beta$ candidate spectrum is, in a sense, an independent measurement
of these contamination levels.  Fixing the background models at some a
priori normalization results in total disregard of the background 
information contained in the measured 2e$^{-}$ spectrum. 
Similarly, allowing the model normalizations to vary freely ignores
the best-estimates.  In addition, the various
background spectra can be qualitatively quite similar (e.g. 
the $^{214}$Bi 2e$^{-}$ sum-energy  spectrum and the theoretical
$\beta\beta$-decay.) These
similarities generate correlated parameters in the fitting
procedure, adding non-physical constraints to the problem.

\begin{table}[t]
\caption{Summary of inputs and results of Extended Maximum Likelihood
fits to the $\beta\beta$-decay 2e$^{-}$ energy spectra.  
All uncertainties and limits represent 90\% confidence levels.}
\renewcommand{\arraystretch}{1.5}
\begin{tabular}{cccr@{}l@{${}\pm{}$}r@{}lr@{}l@{$^{+}_{-}$}l}
$\beta\beta$-decay  & Spectral & Constrained   & 
\multicolumn{4}{c}{Initial}  & \multicolumn{3}{c}{Fitted} \\
Isotope 	    & Model    & by ${\cal L}$ & 
\multicolumn{4}{c}{Estimate} & \multicolumn{3}{c}{Value\tablenotemark[1]}  \\
\tableline   \rule{0mm}{2.75ex}
$^{100}$Mo 
     & $\beta\beta_{2\nu}$     & no  & \multicolumn{4}{c}{---} 
		& 376 & .7 & $^{20.3}_{22.4}$	\\
     & M\"{o}ller  & yes & 20 & .5 & 5 & .9   
		& 19 & .9 & $^{3.9}_{3.9}$ 	\\ 
     & $^{234}$Pa  & no & \multicolumn{4}{c}{---}
		& $\leq$ 15 & \multicolumn{2}{l}{.6}	\\
     & $^{214}$Pb  & yes & 0 & .2 & 3 & .4  
		& $\leq$ 2 & \multicolumn{2}{l}{.3}	\\
     & $^{214}$Bi  & yes & 0 & .12 & 0 & .17 
		& 0 & .11 & $^{0.68}_{0.11}$ 	\\
     & $^{228}$Ac  & yes & 8 & .8 & 5 & .8  
		& 8 & .5 & $^{3.9}_{3.9}$ 	\\    
     & $^{208}$Tl\tablenotemark[2]  & yes & 6 & .5 & 4 & .4
		& 5 & .9 & $^{3.4}_{3.4}$ 	\\   
     & $^{211}$Pb  & no	 & \multicolumn{4}{c}{---}
		& $\leq$ 8 & \multicolumn{2}{l}{.2}	\\   
     & $^{60}$Co   & no	 & \multicolumn{4}{c}{---}
		& $\leq$ 6 & \multicolumn{2}{l}{.3}	\\   
\hline \rule{0mm}{2.75ex}
$^{150}$Nd 
     & $\beta\beta_{2\nu}$     & no  & \multicolumn{4}{c}{---}
		& 414 & .4 & $^{22.3}_{22.4}$	\\
     & M\"{o}ller  & yes & 9 & .2 & 5 & .1   
		& 8 & .9 & $^{3.7}_{3.7}$ 	\\ 
     & $^{234}$Pa  & no & \multicolumn{4}{c}{---}
		& $\leq$ 13 & \multicolumn{2}{l}{.0} 	\\
     & $^{214}$Pb  & yes & 33 & .3 & 16 & .7 
		& 26 & .8 & $^{5.8}_{5.7}$ 	\\
     & $^{214}$Bi  & yes & 9 & .0 & 7 & .7   
		& 9 & .4 & $^{4.5}_{4.5}$ 	\\
     & $^{228}$Ac  & yes & 11 & .9 & 7 & .2 
		& 11 & .2 & $^{4.4}_{4.4}$ 	\\    
     & $^{208}$Tl\tablenotemark[2]  & yes & 8 & .7 & 5 & .5
		& 9 & .0 & $^{3.8}_{3.8}$ 	\\   
     & $^{211}$Pb  & no	 & \multicolumn{4}{c}{---}
		& $\leq$ 5 & \multicolumn{2}{l}{.1} 	\\   
     & $^{60}$Co   & no	 & \multicolumn{4}{c}{---}
		& $\leq$ 3 & \multicolumn{2}{l}{.6}	\\   
     & $^{152}$Eu  & no	 & \multicolumn{4}{c}{---}
		& $\leq$ 5 & \multicolumn{2}{l}{.1}	\\   
\end{tabular}
\label{fitsum}
\tablenotetext[1]{See discussion of uncertainties in Sec.~\ref{ibe}.}
\tablenotetext[2]{Includes $^{212}$Pb and $^{212}$Bi.}
\end{table}

A compromise to these extremes is to suggest the best-estimate values
for the various backgrounds to the fitter as initial values, and then
let the fitter vary the parameters by some amount determined by the
uncertainties associated with each.  This is relatively easy to do in
the Extended Maximum Likelihood scheme, since the energy spectra and
the background estimations are independent experiments.  This means
that the corresponding probabilities may be simply appended to the
likelihood function.  Specifically, the likelihood function is
modified to include the probabilities of the background models taking
on given normalizations as determined by Gaussian probabilities
centered on the best-estimates, with standard deviations equal to the
measured  uncertainties:
\begin{eqnarray*}
{\cal L} = \left[\prod_{i} P(x_{i},\vec{a})\right] e^{-{\cal N}}\:
\left[
\prod_{j}e^{-(a_{j}-\hat{a}_{j})^{2}/2\sigma_{j}^{2}}
\right] ,
\end{eqnarray*}
where the $\hat{a}_{j}$ are the best estimates for the normalizations
of the models, the $\sigma_{j}$ are the uncertainties in those estimates,
and the $a_{j}$ are, as before, the fitted parameters representing the
normalizations of the models.

The estimates summarized in Table~\ref{bgsum}  were used in an EML fit
to the 2e$^{-}$ spectra using the procedure described above.  The
uncertainties in the best estimates were scaled up from the 1-$\sigma$
values quoted in Table~\ref{bgsum} to reflect 90\% confidence levels.
The only unrestricted parameters were those associated with the
$\beta\beta_{2\nu}$ model normalization and the parameters of 
backgrounds not listed in Table~\ref{bgsum}.
The fits were performed on
the standard spectra (i.e.\ no cosine or energy cuts), and consisted
of a fit to all events in the sum-energy and singles-energy spectra. 
Table~\ref{fitsum} summarizes the inputs and results of the fitting
procedure.

The uncertainties in the fitted normalizations are calculated by the
MINOS routine within the Minuit minimization package, and are derived 
by considering the shape of the likelihood function near the maximum.
This procedure underestimates the uncertainties of the 
fitted components in Table~\ref{fitsum}
because of the use of both sum 
and singles spectra in the fits.  MINOS thinks it is working with 
N + 2N = 3N independent events, but the 2N electrons from the singles 
spectrum provide no additional information for the overall normalization.
We re-estimate the uncertainty in the number of $\beta\beta$ 
events by repeating the fit on the sum spectrum only.  The resulting 
$\beta\beta$ component in this case is similar, so 
the sum-spectrum uncertainties for N$_{\beta\beta}$ are 
used in Table~\ref{hlparmtab} and the 
half-life calculations.

A second fit was performed on each data sample by removing 
models for which only an upper limit
was obtained by the first fit.
The fitted normalizations of the other models were found to be 
unaffected by this change. 

Figures~\ref{mofitspc}
and~\ref{ndfitspc} show the fitted model spectra superimposed on the
experimental data. 
As described in Sec.~\ref{emlintro}, the fit
considers each event independently; the binning shown in
Figs.~\ref{mofitspc} and~\ref{ndfitspc} is for illustrative purposes
only.
                                                                
\subsubsection{Systematic Uncertainties in EML Fit}

The EML fit is a point-by-point fit, rather than a fit to a binned
histogram; this feature makes it simple to estimate the effect of the
TPC's energy resolution on the fitted results.  The energy of every
electron in the data set has an uncertainty associated with it, so by
randomly altering each of these energies in keeping with a Gaussian
distribution defined by the best-estimate energy and its uncertainty,
a hypothetical energy spectrum can be generated from the altered
experimental data. This new spectrum can then be analyzed with the EML
fitter, and will probably produce results slightly different from the
unaltered data; these differences are due solely to the energy
resolution of the experiment, not the statistics of the data set. 
This procedure was repeated about 1000 times on both the $^{100}$Mo
and $^{150}$Nd data sets, and distributions of $\beta\beta_{2\nu}$
fits were generated. The widths of the distributions were quite small,
and imply an uncertainty of only about 0.5\%.

\subsection{Half-Life Calculations}
\label{halflifecalc}

The half-life of the $\beta\beta$-decay isotope under investigation is
given by the fundamental quantities determined in this experiment as
\begin{equation}           
T_{1/2} = \left(\frac{\epsilon_{\beta\beta}}{100}\right) \: 
                   f_{\rm{enr}} \: f_{\rm{mol}} 
         \left(\frac{M_{\rm{tot}}}{W_{\rm{mol}}}\right) 
                   N_{\rm{A}} 
         \left(\frac{t_{\rm{live}}}{N_{\beta\beta}}\right)
                   \ln 2 .
\label{hleq2}
\end{equation}

\begin{table}
\caption{Parameters used in half-life
determination.  The ``Source mass'' is the Mo or Nd$_{2}$O$_{3}$ material 
actually used, not just the $\beta\beta$-decay atoms.  ``Observed 
$\beta\beta$-decay events'' is based on maximum likelihood fit to data;
see text for description of other methods and results.  Uncertainties 
are $1\sigma$. }
\begin{tabular}{clcc}
Param-       &             & \multicolumn{2}{c}{Parameter Value} \\ \cline{3-4}
\vspace{1mm}
eter  & Description     & $^{100}$Mo Run     & $^{150}$Nd Run \\
\tableline   \rule{0mm}{2.75ex}
$\varepsilon_{\beta\beta}$ \tablenotemark[1] & 2e$^{-}$ efficiency (\%) & 
$10.1\pm 0.1$ & $11.5\pm 0.1$ \\
$f_{\rm{enr}}$ & Isotope enrichment          & $0.974$           &  $0.91$  \\
$f_{\rm{mol}}$ & $\beta\beta$ atoms/molecule &
  $1$              & $2$              \\
$M_{\rm{tot}}$ & Source mass (g)  & $16.7\pm 0.1$    &  $15.5\pm 0.1$ \\
$W_{\rm{mol}}$ & Mol.\ wt.\ (g/mol) & $99.91$          &  $346.4$   \\
$N_{\rm{A}}$   & Avogadro \# (mol$^{-1}$) &
\multicolumn{2}{c}{$6.022\times10^{23}$} \\
$t_{\rm{live}}$& Live-time (h) & $3275\pm 2$  &   $6287\pm 3$  \\
$N_{\beta\beta}$ 	   & Observed $\beta\beta$ events & 
  $377^{+21}_{-29} \pm 2 $  &  $414^{+23}_{-25} \pm 2 $ 
\end{tabular}
\label{hlparmtab}
\tablenotetext[1]{10~\% systematic uncertainty not included.}
\end{table}

The parameters in Eq.~(\ref{hleq2}) are described in
Table~\ref{hlparmtab}, which also summarizes their values.
Using these values,
the EML-based half-lives for the two isotopes studied are
\begin{eqnarray*}
T_{1/2}^{2\nu}(^{100}\rm{Mo}) &=& (6.82^{+0.38}_{-0.53}\pm 0.68)\times
     10^{18}\;\rm{y}, \\
T_{1/2}^{2\nu}(^{150}\rm{Nd}) &=& (6.75^{+0.37}_{-0.42}\pm 0.68)\times
     10^{18}\;\rm{y}. 
\end{eqnarray*}
A comparison with other experiments can be found in 
Table~\ref{compresults}, and with theory, in Table~\ref{theorytab}.

\begin{table}
\caption{Comparison of the results produced by this work to those from
other experiments.  Uncertainties and half-life limits
are at the 90\% (68\%) confidence levels.  Limits and uncertainties in
square braces indicate that the confidence level was not specified.}
\renewcommand{\arraystretch}{1.5}
\begin{tabular}{r@{}l@{ }l@{ }l@{${}\pm{}$}r@{}lr@{ }r@{}lr@{ }r@{}ll}
	 \multicolumn{6}{c}{$T_{1/2}^{2\nu}$}  
	& \multicolumn{3}{c}{$T_{1/2}^{0\nu}$} 
        & \multicolumn{3}{c}{$T_{1/2}^{0\nu,\chi}$}  & \\
 	\multicolumn{6}{c}{($10^{18}$ y)}  
	& \multicolumn{3}{c}{($10^{21}$ y)} 
	& \multicolumn{3}{c}{($10^{20}$ y)} & Group \\
\tableline  
\multicolumn{13}{c}{$^{100}$Mo Run}    \\
\hline 
	 6 & .82 & $^{+}_{-}$ & $^{(0.38)}_{(0.53)}$ & (0 & .68)
	& $>$ & 1 & .23 
	& $>$ & 3 & .31
	& This work \\
	 11 & .6 & $^{+}_{-}$ & \multicolumn{3}{l}{$^{(3.4)}_{(0.8)}$}
	& \multicolumn{3}{c}{---}
	& \multicolumn{3}{c}{---}
	& UCI~\cite{ucibb90} \\
	 \multicolumn{6}{c}{---}
	& $>$ & 2 & .16
	& $>$ & 3 & .9 
	& UCI~\cite{erice} \\
         11 & .5 & $^{+}_{-}$ & \multicolumn{3}{l}{$^{5.4 (3.0)}_{2.8 (2.0)}$}
	& $>$ & 2 & .6
	& \multicolumn{3}{c}{---}
	& Osaka~\cite{ejiri91a,ejiri91b} \\
	 3 & .3 & $^{+}_{-}$ & \multicolumn{3}{l}{$^{(2.0)}_{(1.0)}$}
	& $>$ & (0 & .71) 
	& \multicolumn{3}{c}{---}
	& INR~\cite{vasilev} \\
	 10 & & $\pm$ & [0.8] & [2 & ]
	& \multicolumn{3}{c}{---}
	& \multicolumn{3}{c}{---}
	& NEMO~\cite{NEMO} \\
	 9 & .5 & $\pm$ & [0.4] & [0 & .9] 
	& $>$ & 6 & .4
	& $>$ & 5 & 
	& NEMO 2~\cite{NEMO2} \\
\hline 
\multicolumn{13}{c}{$^{150}$Nd Run} \\    
\hline 
	 6 & .75 & $^{+}_{-}$ & $^{(0.37)}_{(0.42)}$ & (0 & .68) 
	& $>$ & 1 & .22                    
        & $>$ & 2 & .82
	& This work \\
         \multicolumn{6}{c}{---}  
	& $>$ & 2 & .1 
        & $>$ & 5 & .3
	& UCI~\cite{erice} \\
         17 & & $^{+}_{-}$ & $^{10}_{5}$ & 3 & .5
	& \multicolumn{3}{c}{---}
	& \multicolumn{3}{c}{---}
	& ITEP/INR~\cite{artemiev} \\
\end{tabular}
\label{compresults}
\end{table}

\begin{figure}
\vskip -0.5 cm
\mbox{\epsfxsize=8.6 cm\epsffile{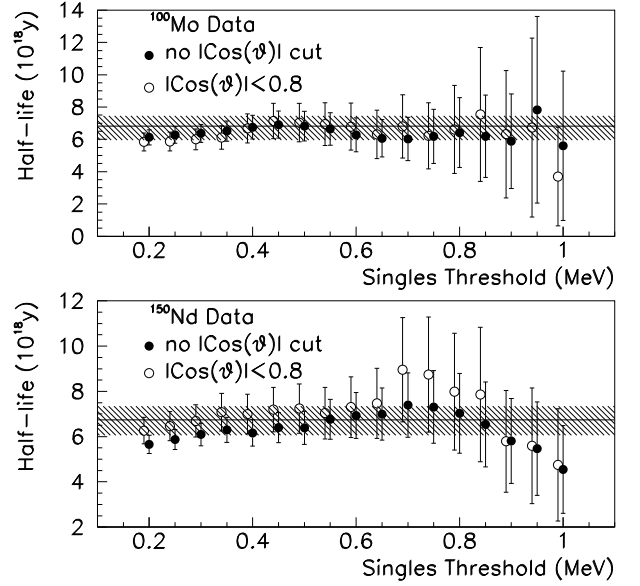}}
\caption{Graphical representation of half-lives derived from the number of 
events at various singles energy thresholds. 
(Open circles are shifted by \mbox{-0.01~MeV} for clarity.)
The hatched band represents the result of the the EML fit to the energy 
spectra. 
Uncertainties are statistical, and represent 90\%
confidence levels; systematic uncertainties contribute approximately
10\% in addition to those shown.  (One should keep in mind that nearby 
points are strongly correlated.)}
\label{cutlifeplot}
\end{figure}

{\label{singcuthl}}
The estimated number of $\beta\beta$-decay events based on
single-electron energy threshold cuts may also be used in lieu of the
EML-based estimates in the half-life calculations. The half-lives
derived from those values are plotted in Fig.~\ref{cutlifeplot}. The
hatched regions in the figure represent the result of the EML
half-life calculation.  The EML result agrees nicely with the the
singles-energy threshold method near the favored threshold of about
500~keV. This
figure does not include systematic errors, which contribute
approximately the same uncertainty to each measurement.

\begin{table}
\caption{Theoretical half-lives for $^{100}$Mo and $^{150}$Nd. 
Zero-neutrino predictions assume $<\!m_{\nu}\!>\,= 1$~eV. }
\begin{tabular}{cr@{}l@{${}\times{}$}lr@{}l@{${}\times{}$}lc}
Isotope & \multicolumn{3}{c}{$T^{2\nu}_{1/2}$ (y)} 
        & \multicolumn{3}{c}{$T^{0\nu}_{1/2}$ (y)} & Ref. \\
\tableline  
$^{100}$Mo    & 1 & .13 & $10^{18}$ & 1 & .27 & $10^{24}$ &
          \cite{epl90} \\
        & $2.87\!-\!7$ & .66 & $10^{18}$ & $2.6\!-\!4$ & .7 & $10^{23}$ &
	  \cite{suhonen94} \\
        & 6 & .0 & $10^{18}$ & 1 & .9 & $10^{24}$ & 
          \cite{engel88} \\
\hline 
$^{150}$Nd    & 7 & .37 & $10^{18}$ & 3 & .37 & $10^{22}$ & 
          \cite{epl90} \\
        & \multicolumn{3}{c}{---} & 1 & .05 & $10^{24}$ & \cite{hirsch95a} \\
	& 6 & .73 & $10^{18}$ & \multicolumn{3}{c}{---} & \cite{hirsch95b} \\ 
\end{tabular}
\label{theorytab}
\end{table}

\subsection{Kurie Plots}

\begin{figure}
\vskip -0.5 cm
\mbox{\epsfxsize=8.6 cm\epsffile{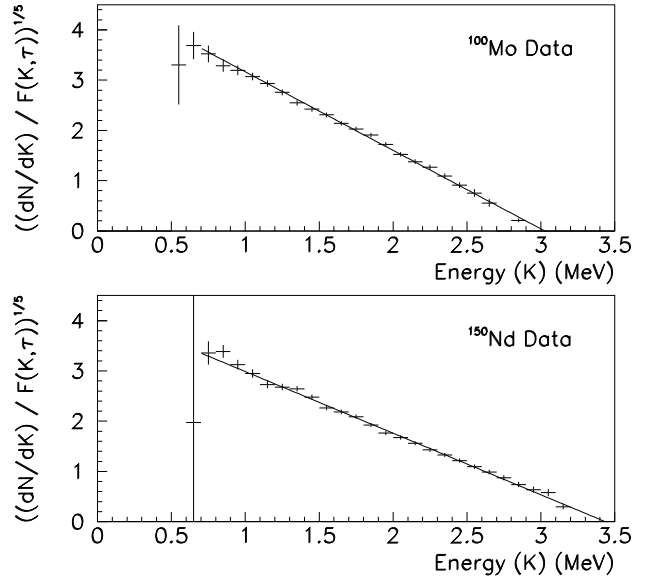}}
\caption{The Kurie plots, for $^{100}$Mo and $^{150}$Nd.  Also shown are 
linear fits to the points and the x-intercept indicating the end-point 
energy.($F(K,\tau)=(K-2\tau)[f_0(K)+f_{\tau}(K)]$. See text.)}
\label{kurie}
\end{figure}

The Kurie plots were generated by first
subtracting the background 2e$^{-}$ sum energy spectra, from the 
candidate spectrum, using the amounts given by the 
EML fit results.
The next step involved obtaining the true 2e$^{-}$ residual spectrum. 
Here Eq.~(\ref{truespec}) was used 
where K and K$^{\prime}$ now refer to the 
true and measured sum energy and 
\begin{eqnarray*}
g(K) & = & \frac{N^{\rm mc}_{\rm shift}(K)}{N_{\tau}(K)},
\end{eqnarray*}
where N$_{\tau}$(K)
is the true number of events in the sum energy spectrum with
singles energies greater than 0.25~MeV.  
Also, for this calculation, $P(K,K^{\prime})$ 
and $N^{\rm mc}_{\rm shift}(K)$ were determined using the theoretical 
$\beta\beta$ spectrum as input to the Monte Carlo. (This was later 
replaced by a randomly generated spectrum in order to determine the 
systematic uncertainty.)
Finally the Kurie plot values were calculated from the true residual spectrum 
using the formula~\cite{calcium} 
\begin{eqnarray*}
(Q_{\beta\beta} - K) & \propto & ((dN/dK)/\{(K-2\tau) 
[f_0(K)+f_{\tau}(K)]\})^{1/5}  
\end{eqnarray*}
where
\begin{eqnarray*}
f_0(K) & = & K^4/30 + K^3m/3 + 4K^2m^2/3 + 2Km^3 + m^4, \\
f_ {\tau}(K) & = & \tau(K - \tau)[K^2/15 + 2Km/3 + 2m^2/3 
\nonumber \\
& & \mbox{} + \tau(K - \tau)/5],
\end{eqnarray*}
and $m$ is the electron mass.
This formula was obtained by including the singles threshold, 
$\tau$, and using the Primakoff-Rosen approximation
for the Coulomb effect; the approximation produces 
$<\!\!1$\% distortion over the plotted range.
The Kurie plot (see Fig.~\ref{kurie}) projects to 
an end-point energy for the $^{100}$Mo source of
($3.03\pm0.02\pm0.06$)~MeV, in
good agreement with the mass difference of 3.03~MeV.  The 
corresponding $^{150}$Nd value is ($3.44\pm0.02\pm0.06$)~MeV, 
also in agreement with the mass difference of 3.37~MeV.

\subsection{Exotic Decay Limits}
                                                    
\subsubsection{$\beta\beta_{0\nu}$-decay Half-Life Limits}

The expected $\beta\beta_{0\nu}$ distributions were generated with a
Monte Carlo simulation. For $^{100}$Mo, the FWHM lies between 2.89 and 3.18~MeV
and for $^{150}$Nd, this region is bounded by 3.21 and 3.54~MeV.
There were no counts from either isotope observed within the cut
regions, and poisson statistics allows us to reject 2.3 events at the 
90\% confidence level.  

Half-life limits are derived from the above limits by applying
Eq.~(\ref{hleq2}).  Two of the parameters listed in
Table~\ref{hlparmtab} must be altered for this calculation:  The
90\%~CL decay limits are used in place of the number of observed
decays, and a new efficiency for the $\beta\beta_{0\nu}$ case
is given by Monte Carlo as 
$\epsilon_{\beta\beta}=11.12\pm0.07$ for $^{100}$Mo and 
$\epsilon_{\beta\beta}=11.50\pm0.07$ for $^{150}$Nd.  
With these parameters, the half-life limits are
(at 90\%~CL)
\begin{eqnarray*}
T_{1/2}^{0\nu}(^{100}\rm{Mo}) &>& 1.23 \times 10^{21}\;\rm{y}, 
\\
T_{1/2}^{0\nu}(^{150}\rm{Nd}) &>& 1.22 \times 10^{21}\;\rm{y}.
\end{eqnarray*}

(See Table~\ref{theorytab} for theoretical predictions.)
These $\beta\beta_{0\nu}$ half-lives do not give competitive limits 
on neutrino mass and right-handed current parameters.

\subsubsection{$\beta\beta_{0\nu,\chi}$-decay Half-Life Limits}

The expected energy spectra for the $\beta\beta_{0\nu,\chi}$
mode are continuous distributions, spanning the entire energy range
available to the $\beta\beta_{2\nu}$ channel.  As such, limits
on the contribution from  $\beta\beta_{0\nu,\chi}$ cannot be
determined by simply counting events in a predetermined region, and
the maximum-likelihood method described in Sec.~\ref{emlintro} must
be used.  We restrict our analysis to the ``ordinary majoron'', index=1 in
ref.~\cite{burgess3}.

Maximum likelihood fits were performed on both the $^{100}$Mo and
$^{150}$Nd data sets, using the same fitting program previously used
for the $\beta\beta_{2\nu}$ analysis.  In this case, the fit was
restricted to energies well above the majority of $\beta\beta_{2\nu}$
backgrounds, where the majoron decay signature is significant (i.e.\
singles-energies greater than 750~keV, and sum-energies greater than
1.5~MeV).  The fit was performed jointly over both the sum- and
singles-energy spectra, and included only two models:  the
$\beta\beta_{2\nu}$ and $\beta\beta_{0\nu,\chi}$ Monte
Carlo energy spectra.
Under the assumption that poor energy resolution at high energies will
adversely effect the reliability of our $\beta\beta_{0\nu,\chi}$
limit estimates, we cut the data at $|\cos\vartheta|\!<\!0.8$.

In both the $^{100}$Mo and $^{150}$Nd cosine-cut spectra, the EML
fitting procedure indicates that the data sample is consistent with
there being no $\beta\beta_{0\nu,\chi}$ events in either
spectrum.  The algorithm used by Minuit to determine confidence
intervals, MINOS, cannot be used when the most likely solution is
against a physical boundary (in this case, negative normalizations are
non-physical solutions), so a direct integration of the likelihood
function must be performed in order to calculate the confidence
intervals.  

The 90\% confidence limit on the number of
$\beta\beta_{0\nu,\chi}$ events in our data sets 
is 7.76 events in the $^{100}$Mo data set and
9.73 events in the $^{150}$Nd data set.  Half-life limits can be
calculated by using Eq.~(\ref{hleq2}) with these decay limits in
place of an observed number of decays, and using the appropriate 
$\beta\beta_{0\nu,\chi}$ efficiencies of
$\epsilon_{\beta\beta} = 10.13\pm0.06$ for $^{100}$Mo and 
$\epsilon_{\beta\beta} = 11.25\pm0.07$ for $^{150}$Nd, giving       
\begin{eqnarray*}                          
T_{1/2}^{0\nu,\chi}(^{100}\rm{Mo}) &>& 3.31 \times 10^{20}\;\rm{y}, 
\\
T_{1/2}^{0\nu,\chi}(^{150}\rm{Nd}) &>& 2.82 \times 10^{20}\;\rm{y}, 
\end{eqnarray*}
at the 90\% confidence limit.  
The $\beta\beta_{0\nu,\chi}$ half-life limit published by our group
in 1994~\cite{erice} was
based on an analysis of binned data, and the fitting routine has since
improved.  We consider the current analysis to be more reliable.
We use the matrix elements of Ref.~\cite{epl90}
 to put
limits on the effective neutrino-majoron coupling constant and obtain
\begin{eqnarray*}                          
<\!g_{\nu,\chi}\!>(^{100}\rm{Mo}) & < & 6.26 \times 10^{-4}, 
\\
<\!g_{\nu,\chi}\!>(^{150}\rm{Nd}) & < & 9.96 \times 10^{-5},
\end{eqnarray*}
at the 90\% confidence level.
                                              

\section{Conclusions}

Although our choice of a very thin source plane precluded sufficient 
source mass to give competitive limits on $\beta\beta_{0\nu}$, 
it did result in exceptionally clean $\beta\beta_{2\nu}$ energy spectra 
composed overwhelmingly of $\beta\beta$-decay events.  These spectra were 
produced without subtraction of any untagged background events.
The derived Kurie plots are straight over a broad energy range, and 
intercept the energy axis close to the expected Q$_{\beta\beta}$ 
values.

Improvements in fitting techniques and in determination of the detector 
efficiency have resulted in a {$^{100}$Mo}
half-life  that is shorter
than we reported previously~\cite{ucibb90}.
We now believe that the maximum-likelihood 
fit performed
on that early data set significantly over-estimated the background
contamination, contributing to an artificially high half-life estimate. 

The intriguing high-energy anomalies in
our earlier $\beta\beta$-decay sum-energy spectra 
did not withstand the improved energy resolution that resulted 
from a doubled magnetic field strength.  No suggestion of 
$\beta\beta_{0\nu,\chi}$ remains in either sum spectrum.  At our level 
of sensitivity, the double beta decay phenomenon is well described by 
standard theory.

\acknowledgments

We express our thanks for the contributions of Steve Elliott, Alan Hahn, 
and the many students at UCI who have had a hand in this effort.
We thank Ben Wilkinson, Project Manager at the Hoover Dam, 
and his successor, Blaine Hamann, for their enthusiastic 
provision of an underground site for this experiment, and we are 
grateful to Bill Sharp for logistical support at the dam.
This work was supported by the U.S. Department of Energy under GRANT 
DE-FG03-91ER40679.


\end{document}